\documentclass[aps,pra,longbibliography,superscriptaddress,nofootinbib,twocolumn]{revtex4-1}

\usepackage[T1]{fontenc}
\usepackage[utf8]{inputenc}

\usepackage{float}
\usepackage{graphicx}
\usepackage{epsfig,epstopdf}
\usepackage[table]{xcolor}

\usepackage[english]{babel}

\PassOptionsToPackage{hyphens}{url}
\usepackage[hyphenbreaks]{breakurl}

\usepackage{times}
\usepackage{booktabs}

\usepackage{amsmath}
\usepackage{amsfonts}
\usepackage{amssymb}
\usepackage{amsthm}
\usepackage{bbm}
\usepackage{mathtools}
\usepackage{siunitx}

\usepackage{MnSymbol}
\usepackage{adjustbox}
\usepackage{multirow}

\usepackage{enumerate}

\usepackage{tikz}
\usetikzlibrary{calc}
\usetikzlibrary{positioning}

\usepackage{algorithm}
\usepackage{algpseudocode}
\algrenewcommand\algorithmicrequire{\textbf{Input:}}
\algrenewcommand\algorithmicensure{\textbf{Output:}}

\PassOptionsToPackage{pdftex,hyperfootnotes=false,pdfpagelabels}{hyperref}
\usepackage{hyperref}
\hypersetup{%
    colorlinks=true, linktocpage=true, pdfstartpage=3, pdfstartview=FitV,%
    breaklinks=true, pdfpagemode=UseNone, pageanchor=true, pdfpagemode=UseOutlines,%
    plainpages=false, bookmarksnumbered, bookmarksopen=true, bookmarksopenlevel=1,%
    hypertexnames=true, pdfhighlight=/O,
    urlcolor=blue!60!black, linkcolor=blue!60!black, citecolor=gray, 
}   

\usepackage[capitalise,compress]{cleveref}

\makeatletter 
\hypersetup{
      pdftitle = {
            Robustly learning the Hamiltonian dynamics of a superconducting quantum processor
            },
      pdfauthor = {
            Dominik Hangleiter, Ingo Roth, Jonas Fuksa, Jens Eisert, Pedram Roushan
            },
	     pdfsubject = {
            quantum simulation, 
            quantum system identification}
            ,
	     pdfkeywords = {
            Hamiltonian tomography, 
            Sycamore, 
            quantum advantage, 
            superconducting qubits, 
            bosons,
            super-resolution,
            geometric optimization, 
            state preparation and measurement error,
        }
	    }
\makeatother

\newtheorem*{problem*}{Problem}

\newcommand{\myleft}{\mathopen{}\mathclose\bgroup\left}
\newcommand{\myright}{\aftergroup\egroup\right}
\newcommand{\e}{\ensuremath\mathrm{e}}
\renewcommand{\i}{\ensuremath\mathrm{i}}
\newcommand{\id}{\ensuremath\mathbbm{1}}

\DeclareMathOperator{\Tr}{Tr}
\renewcommand{\Re}{\operatorname{Re}}
\renewcommand{\Im}{\operatorname{Im}}

\DeclareMathOperator{\rank}{rank}

\DeclareMathOperator{\Id}{Id}

\DeclareMathOperator{\diag}{diag}

\DeclareMathOperator{\sign}{sign}

\renewcommand{\O}{\operatorname{O}}

\DeclareMathOperator\tr{Tr}

\DeclareMathOperator{\Hankel}{Hk}

\makeatletter 
\newcommand\complexity@possiblymakesmaller[1]{#1} 
\newcommand\complexity@fontcommand{\mathsf} 
\newcommand{\ComplexityFont}[1]{%
{\ensuremath{\complexity@possiblymakesmaller{\complexity@fontcommand{#1}}}}
}
\makeatother

\newcommand{\CC}{\mathbb{C}}
\newcommand{\RR}{\mathbb{R}}

\newcommand{\mc}[1]{\mathcal{#1}}

\newcommand{\norm}[1]{\left\Vert #1 \right\Vert} 

\newcommand{\ket}[1]{\left.\left|{#1}\right.\right\rangle}

\newcommand{\bra}[1]{\left.\left\langle{#1}\right.\right|}

\newcommand{\braket}[2]{\left\langle #1 \middle| #2 \right\rangle}

\newcommand{\ketbra}[2]{\ket{#1} \!\! \bra{#2}}

\newcommand{\proj}[1]{\ketbra{#1}{#1}}
\newcommand{\sandwich}[3]
  {\left\langle  #1 \right| #2 \left| #3 \right\rangle}

\newcommand{\A}{\mc{A}} 

\newcommand{\eig}{\mathrm{eig}}

\usepackage{bibunits}
\defaultbibliography{hamiltonian_tomography}
\defaultbibliographystyle{./myapsrev4-1}

\usepackage{comment}

\definecolor{ingo}{rgb}{.1,.8,.2}

\definecolor{jens}{rgb}{0,.8,.6}

\definecolor{pedram}{rgb}{0.75,0.5,0.05}

\definecolor{dominik}{rgb}{0.4,.0,0.6}

\definecolor{jonas}{rgb}{.4,0,.7}

\newcommand{\fu}{Dahlem Center for Complex Quantum Systems, Freie Universit{\"a}t Berlin, 14195 Berlin, Germany}

\newcommand{\hzb}{Helmholtz-Zentrum Berlin f{\"u}r Materialien und Energie, 14109 Berlin, Germany}

\newcommand{\quics}{Joint Center for Quantum Information and Computer Science (QuICS), University of Maryland and NIST, College Park, MD 20742, USA}
\newcommand{\jqi}{Joint Quantum Institute (JQI), University of Maryland and NIST, College Park, MD 20742, USA}
\newcommand{\tii}{Quantum Research Center, Technology Innovation Institute (TII), Abu Dhabi}

\newcommand{\google}{Google Quantum AI, Mountain View, CA, USA}
\begin{document} 

\title{
Robustly learning the Hamiltonian dynamics of a superconducting quantum processor
}

\author{Dominik Hangleiter}
\email[D.H. and I.R. have contributed equally and correspond @ ]{mail@dhangleiter.eu, ingo.roth@tii.ae}
\affiliation{\quics}
\affiliation{\jqi}

\author{Ingo Roth}
\email[D.H. and I.R. have contributed equally and correspond @ ]{mail@dhangleiter.eu, ingo.roth@tii.ae}
\affiliation{\tii}
\affiliation{\fu}

\author{Jon\'a\v{s} Fuksa}
\affiliation{\fu}

\author{Jens Eisert}
\affiliation{\fu}
\affiliation{\hzb}

\author{Pedram Roushan}
\affiliation{\google}





\begin{abstract}
The required precision to perform quantum simulations beyond the capabilities of classical computers imposes major experimental and theoretical challenges. The key to solving these issues are precise means of characterizing analog quantum simulators. Here, we robustly estimate the free Hamiltonian parameters of bosonic excitations in a superconducting-qubit analog quantum simulator from measured time-series of single-mode canonical coordinates. 
We achieve high levels of precision in estimating the Hamiltonian parameters by exploiting a priori knowledge, making it robust against noise and \emph{state-preparation and measurement} (SPAM) errors. Importantly, we are also able to obtain tomographic information about those SPAM errors from the same data, crucial for the experimental applicability of Hamiltonian learning in dynamical quantum-quench experiments. Our learning algorithm is scalable both in terms of the required amounts of data and post-processing. To achieve this, we develop a new super-resolution technique coined tensorESPRIT for frequency extraction from matrix time-series. The algorithm then combines tensorESPRIT with constrained manifold optimization for the eigenspace reconstruction with pre- and post-processing stages. For up to 14 coupled superconducting qubits on two Sycamore processors, we identify the Hamiltonian parameters---verifying the implementation on one of them up to sub-MHz precision---and construct a spatial implementation error map for a grid of 27 qubits. Our results constitute an accurate implementation of a dynamical quantum simulation that is characterized using a new diagnostic toolkit for understanding, calibrating, and improving analog quantum processors.
\end{abstract}
\maketitle
\begin{bibunit}

Analog quantum simulators promise to shed light on fundamental questions of physics that have remained elusive to the standard methods of inference~\cite{feynman_simulating_1982,Lloyd}. 
Recently, enormous progress in controlling individual quantum degrees of freedom has been made towards making this vision a reality \cite{BlochSimulation,BlattSimulator,MonroeTimeCrystal,ebadi_quantum_2021}.
While in digital quantum computers small errors can be corrected~\cite{aharonov_fault-tolerant_2008}, it is intrinsically difficult to error-correct analog devices. 
Yet, the usefulness of analog quantum simulators as computational tools depends on the error of the implemented dynamics. 
Meeting this requirement hinges on devising characterization methods that not only yield a benchmark of
the overall functioning of the device~\cite[e.g.,][]{derbyshire_randomized_2020,shaffer_practical_2021,helsen_matchgate_2020}, but more importantly provide \emph{diagnostic information} about the sources of errors. 

Developing characterization tools for analog quantum simulators requires hardware developments as well as theoretical analysis and method development. 
With the advent of highly controlled quantum systems, efficient methods for identifying certain Hamiltonian parameters from dynamical data have been devised for specific classes of Hamiltonians.
Key ideas are the use of Fourier analysis~\cite{schirmer_experimental_2004,cole_identifying_2005,cole_identifying_2006,cole_precision_2006,schirmer_physics-based_2008,schirmer_two-qubit_2009,oi_quantum_2012} and tracking the dynamics of single excitations~\cite{burgarth_coupling_2009,burgarth_indirect_2011,burgarth_indirect_2009,di_franco_hamiltonian_2009,wiesniak_finding_2010,burgarth_quantum_2012}. 
For general Hamiltonian models, specific algebraic structures of the Hamiltonian terms can be exploited~\cite{zhang_quantum_2014,sone_exact_2017}. 
Generalizing these ideas, a local Hamiltonian can be learned from a single eigenstate or its steady state~\cite{garrison_does_2018,qi_determining_2019,chertkov_computational_2018,bairey_learning_2019,bairey_learning_2020,evans_scalable_2019} or using quantum-quenches \cite{li_hamiltonian_2020,czerwinski_hamiltonian_2021}, an approach dubbed `correlation matrix method' \cite{elben_randomized_2023}.
Alternatively, one can apply general-purpose machine-learning methods~\cite{valenti_hamiltonian_2019,bienias_meta_2021,krastanov_stochastic_2019,che_learning_2021,wilde_scalably_2022}.
More recently, optimal theoretical guarantees have been derived for Hamiltonian learning schemes \cite{yu_robust_2023-1,huang_learning_2023,li_heisenberg-limited_2023} based on Pauli noise tomography \cite{flammia_efficient_2020,harper_efficient_2020}.
Crucially, these protocols assume perfect mid-circuit quenches, which---as we find here---can be a limiting assumption in practice.

This recent rapid theoretical development is not quite matched by concomitant experimental efforts.
The effectiveness of some of these methods has been demonstrated for the estimation of a small number of coupling parameters of fixed two- and three-qubit Hamiltonians in \emph{nuclear magnetic resonance} (NMR) experiments~\cite{lapasar_estimation_2012,hou_experimental_2017,chen_experimental_2021,zhao_characterizing_2021}.
While in NMR, the dominant noise process is decoherence, in tunable quantum simulators such as superconducting qubits, trapped ions or cold atoms in optical lattices, \emph{state preparation and measurement} (SPAM) errors, as we also demonstrate here, play a central role. 
Initial steps at characterizing such errors as well as the 
dissipative Lindblad 
dynamics for up to two qubits in a superconducting qubit platform have been taken recently \cite{flurin_using_2020,samach_lindblad_2021}. 
Hamiltonian learning of thermal states has recently also been applied in many-body experiments as a means to characterize the entanglement of up to $20$-qubit subsystems whose reduced states are parameterized by the so-called \emph{entanglement Hamiltonian} \cite{kokail_entanglement_2021,kokail_quantum_2021,joshi_exploring_2023}.
The challenge remains to develop and experimentally demonstrate the feasibility of scalable methods for a robust and precise identification of Hamiltonian dynamics of intermediate-size systems subject to both incoherent noise and systematic SPAM errors. 

Here, we develop bespoke protocols to robustly and accurately identify the full Hamiltonian of a large-scale bosonic system and implement those protocols on superconducting quantum processors.
Given the complexity of the learning task, we focus on identifying the non-interacting part of a potentially interacting system.
We are able to estimate the corresponding Hamiltonian parameters as well as SPAM errors pertaining to all individual components of the superconducting chip for up to 14-mode Hamiltonians tuned across a broad parameter regime, in contrast to previous experiments. 
Given the identified Hamiltonians, we quantify their implementation error. 
We demonstrate and verify that a targeted intermediate-size Hamiltonian is implemented on a large region of the superconducting processor with sub-MHz precision in a broad parameter range. 

\begin{figure*}[t]
  \centering
  \includegraphics[width=\textwidth]{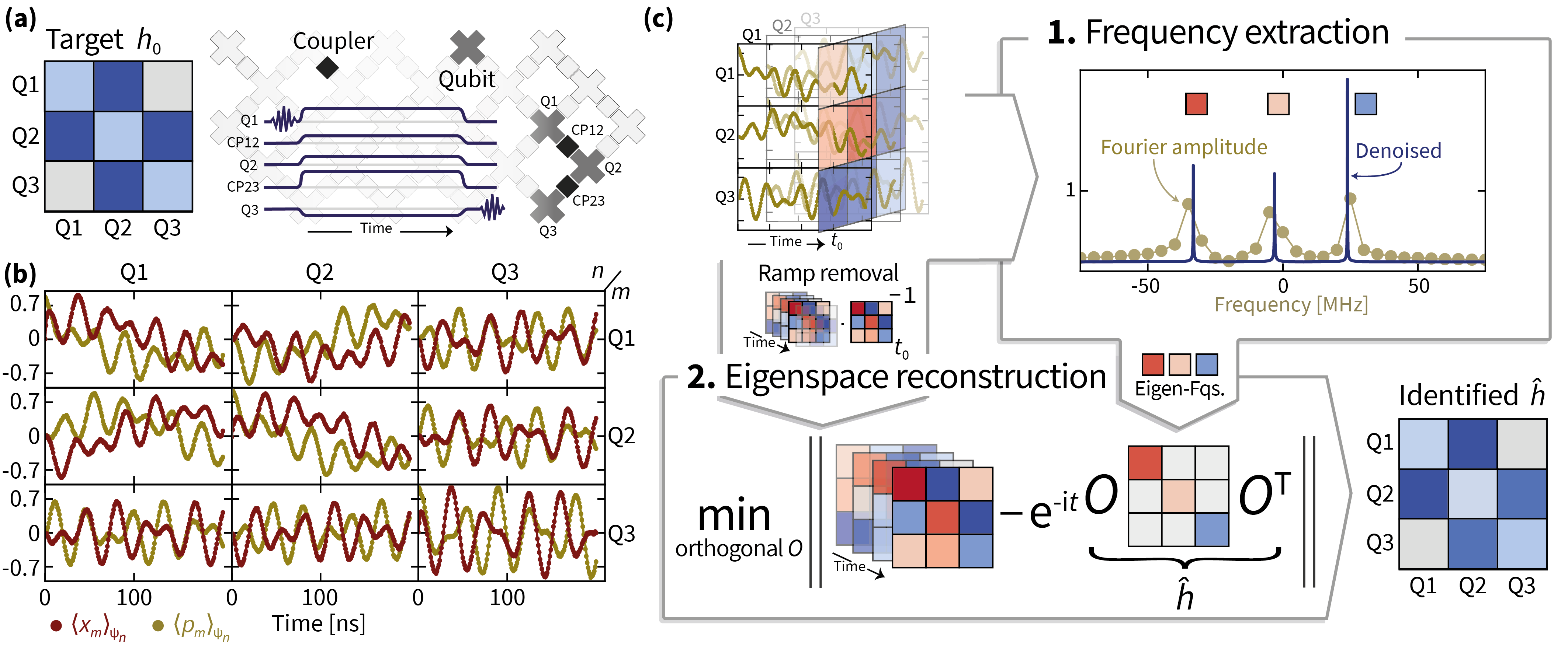}
  \caption{
    \label{fig:outline}
    \textbf{Outline of the experiment and identification algorithm.} \textbf{(a)} The time evolution under a target Hamiltonian $h_0$ is implemented on an part of the Google Sycamore chip (gray) using the pulse sequence depicted in the middle. \textbf{(b)} The expected value of canonical coordinates $x_m$ and $p_m$ for each qubit $m$  over time is estimated from measurements using different $\psi_n$ as input states. \textbf{(c)} The data  shown in (b) for each time $t_0$ can be interpreted as a (complex-valued) matrix with entries indexed by measured and initial excited qubit, $m$ and $n$. The identification algorithm proceeds in two steps: 1.\ From the matrix time-series,
    the Hamiltonian eigenfrequencies are extracted using our newly introduced algorithm coined \emph{tensorESPRIT}, introduced in the SM, or an adapted version of the ESPRIT algorithm. The blue line indicates the denoised, high-resolution signal as `seen' by the algorithm. 2.\ After removing the initial ramp using the data at some fixed time, the Hamiltonian eigenspaces are reconstructed using a non-convex optimization algorithm over the orthogonal group. {We obtain a diagonal orthogonal estimate of the final ramp.} From the extracted frequencies and reconstructed eigenspaces, we can calculate the identified Hamiltonian $\hat h$ that describes the measured time evolution and a tomographic estimate of the initial ramp.} 

\end{figure*}

To this end, building on previous ideas for Hamiltonian identification~\cite{burgarth_indirect_2011,zhang_quantum_2014}, we devise a simple and robust algorithm that exploits the structure of the system at hand.
For the identification we make use of quadratically many experimental time-series tracking excitations via expectation values of canonical coordinates. 
Our structure-enforcing algorithm isolates two core tasks that need to be solved in Hamiltonian identification after suitable pre-processing of the data: 
frequency extraction and eigenspace reconstruction. 

To solve the first task in a robust and structure-specific way, we develop a novel algorithm coined \emph{tensorESPRIT}, which utilizes ideas from super-resolving, denoised Fourier analysis~\cite{RoyPaulrajKailath:1986,Fannjiang:2016:ESPRIT,LiLiaoFannjiang:2019:Super-resolution} and tensor networks to extract frequencies from a matrix time-series. For the second task we use constrained manifold optimization over the orthogonal group~\cite{abrudan_conjugate_2009}.
Crucially, by explicitly exploiting all structure constraints of the identification problem, our method allows us to distinguish and obtain tomographic information about state-preparation and measurement errors.
In the quench-based experiment this information renders identification and verification of the dynamics experimentally feasible in the first place. 
We further support our method development with  numerical simulations of different noise effects and benchmark against more direct algorithmic approaches. 
We find that in contrast to other approaches our method is scalable to larger system sizes out of the reach of our current experimental efforts.

Our work constitutes a detailed case study that lays bare and provides solutions for the difficulties of \emph{practical} Hamiltonian learning in a seemingly simple system. 
It thus provides a blueprint and paves the way for devising practical model-specific 
identification algorithms both for the interaction parameters of bosonic or fermionic systems and more complex settings.
\smallskip

\paragraph*{Setup.}
We characterize the Hamiltonian governing analog dynamics of Google Sycamore chips 
which consist of a two-dimensional array of nearest-neighbour coupled superconducting qubits.
Each physical qubit is a non-linear oscillator with bosonic excitations 
(microwave photons)~\cite{PedramReview}. 
Using the rotating-wave approximation the dynamics governing the excitations of the qubits in the rotating frame can be well described by the Bose-Hubbard Hamiltonian \cite{yanTunableCouplingScheme2018}
\begin{align}
\label{eq:bose-hubbard H}
  H_{\text{BH}} = \sum_i \left( \mu_i a_i^\dagger a_i +\eta_i a_i^\dagger a_i^\dagger a_i a_i\right)- \sum_{ i\neq j} J_{i,j} a_i^\dagger a_j , 
\end{align}
\noindent
where $a^\dagger_i$ and $a_i$ denote bosonic creation and annihilation operators at site $i$, respectively, $\mu \in \mathbb{R}^N$ are the on-site potentials, $J \in \mathbb{R}^{N \times N}$ are the hopping rates between nearest neighbour qubits, and $\eta \in \mathbb{R}^N$ are the strength of on-site interactions. The qubit frequency, the nearest-neighbour coupling between them, and the non-linearity (anharmonicity) set $\mu$, $J$, and $\eta$.
We are able to tune $\mu$ and $J$ on nanosecond timescales, while $\eta$ is fixed for a given setting of $\mu$ and $J$. 
Hence, the Sycamore chip can be used to implement time evolution under Hamiltonians of the form \eqref{eq:bose-hubbard H} at various parameter settings and is therefore an analog simulator.
In a practical application such as in Ref.~\cite{roushan_spectroscopic_2017}, it is crucial to benchmark how accurately the implemented dynamics is described by a targeted Hamiltonian.

Here, we focus on the specific task of identifying the values of $\mu_i$ and $J_{i,j}$. 
The corresponding non-interacting part of the Hamiltonian acting on $N$ modes can be conveniently parametrized as 
\begin{align}
\label{eq:non-interacting h}
 H(h) = - \sum_{ i,j = 1}^N h_{i,j} a_i^\dagger a_j 
\end{align} 
with an $N \times N$ real symmetric parameter matrix $h$ with entries $h_{i,j}$, which is composed of the on-site chemical potentials $\mu_i$ on its diagonal and the hopping energies $ J_{i,j}$ for $i \neq j$. 
The identification of the non-interacting part $H(h)$ of $H_{\text{BH}}$ can be viewed as a first step in a hierarchical procedure for characterizing dynamical quantum simulations with tunable interactions and numbers of particles. 

The non-interacting part $H(h)$ of the Hamiltonian $H_{\text{BH}}$ can be inferred when initially preparing a state where only a single qubit is excited with a single 
photon.
For initial states with a single excitation, the interaction term vanishes, hence effectively $\eta = 0$. Consequently, only the two lowest energy levels of the non-linear oscillators enter the dynamics. 
Therefore, referring to them as qubits (two-level systems) is precise. 
Specifically, we identify the parameters $h_{i,j}$ from dynamical data of the following form. 
We initialize the system in $\ket {\psi_n} \coloneqq  (\id + a_n^\dagger)
\ket{0}^{\otimes N}/{\sqrt 2}$
and measure the canonical coordinates $x_m = (a_m + a_m^\dagger)/2$ and $p_m = (a_m - a_m^\dagger)/(2 \i )$ for all combinations of $m,n = 1, \ldots, N$. 
In terms of the qubit architecture, 
this amounts to local Pauli-$X$ and Pauli-$Y$ basis measurements, respectively. 
We combine the  statistical averages over multiple measurements to obtain an empirical estimator for 
$ \langle a_m(t) \rangle_{\psi_n} = \langle x_m(t) \rangle_{\psi_n}+ \i \,\langle p_m(t) \rangle_{\psi_n} $. 
For particle-number preserving dynamics, this data is of the form 
\begin{align}
    \label{eq:time-evolution data}
    \langle a_m(t) \rangle_{\psi_n}  = \frac 12\exp(-\i t h)_{m,n}\, . 
\end{align}
It therefore directly provides estimates of the entries of the time-evolution unitary at time $t$ in the single-particle sector of the bosonic Fock space.

\begin{figure*}[ht]
  \includegraphics[width = \linewidth]{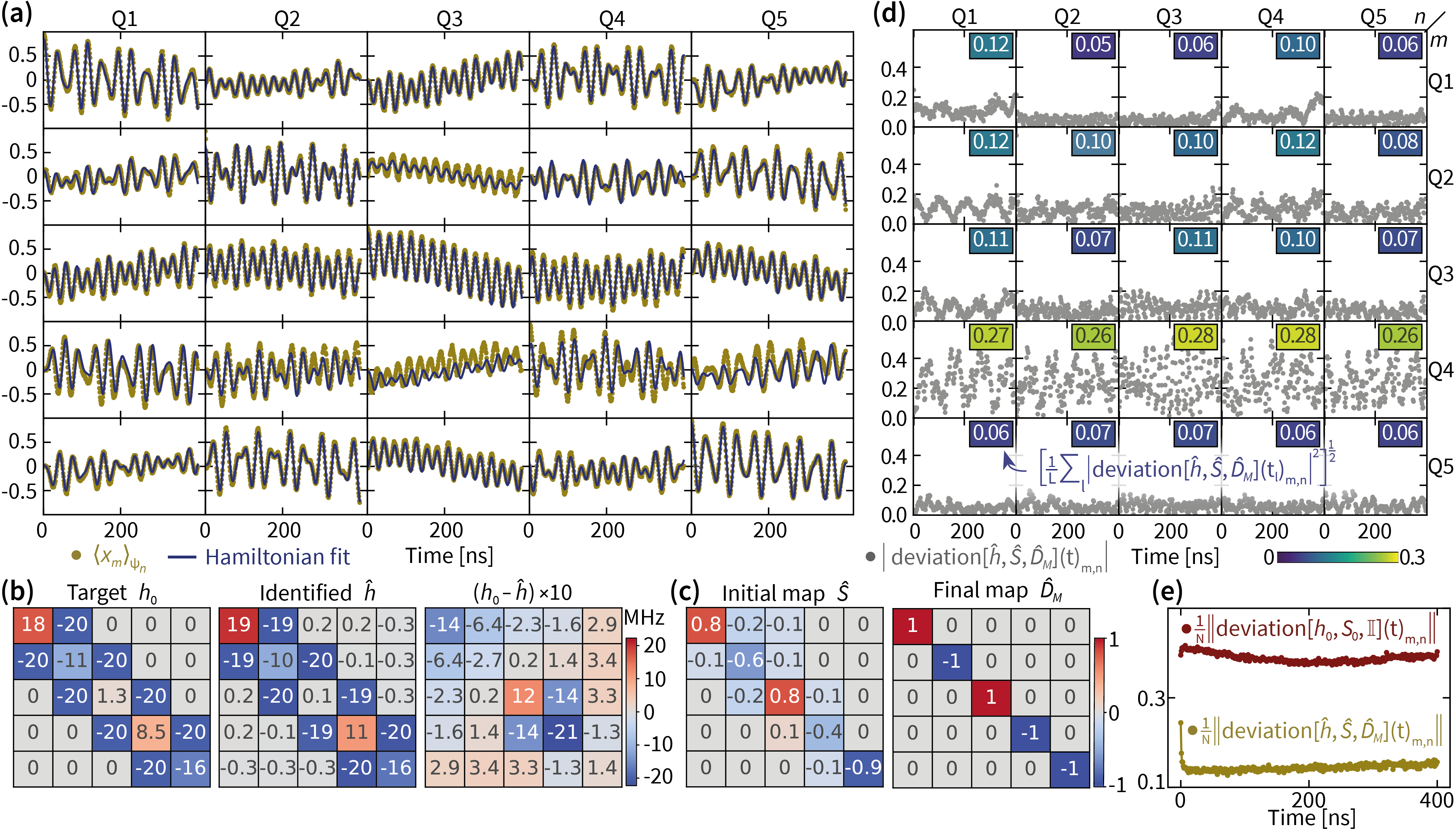}
  \caption{
    \label{fig:single H}
    \textbf{A single Hamiltonian recovery of a 5-mode Hamiltonian and the corresponding time domain data.}
    \textbf{(a)} The full experimental time-series data $\langle x_m(t) \rangle_{\psi_n}$ for $m,n = 1, \ldots, 5$ and the best fit of those data in terms of our model $\frac 12 (M\exp(-\i t h) S)_{m,n}$ for a diagonal and orthogonal $M$ and linear map $S$ (solid lines). 
    \textbf{(b)} The target Hamiltonian matrix $h_0$, the identified Hamiltonian $\hat h$, and the deviation between them. 
    The error of each diagonal entry is
    $\pm (0.16 + 0.99)$ MHz and of each  
    off-diagonal entry 
    $\pm (0.12 + 0.50)$ MHz  
    and comprises of the statistical and the systematic error, respectively. 
    The analog implementation error $\mc E_{\rm analog}(\hat h, h_0)$ is 
    $0.73 \pm (0.07 + 0.62)$\,MHz,  
    and $0.32 \pm  0.00$\,MHz for the eigenfrequencies.
    The analog implementation error $\mc E_{\rm analog}(\hat S,\id)$ of the identified initial map is 
    $0.61 \pm (0.00 + 0.12)$. 
    \textbf{(c)} The real part of the initial map $\hat S$ and the diagonal orthogonal estimate $\hat D_M$ of the final map $M$, inferred from the data using the identified Hamiltonian $\hat h$. 
    \textbf{(d)} Absolute value of the time-domain deviation of the fit from the full experimental data for each time series, given by $\textsf{deviation}[\hat h, \hat S, \hat D_M](t)_{m,n} \coloneqq \langle a_m(t) \rangle_{\psi_n} - \frac 12 \hat D_M \exp(- \i t \hat h) \hat S$. 
    The insets represent the root-mean-square deviation of the Hamiltonian fit from the experimental data per time series,
    averaged over the evolution time for each matrix entry $(m,n)$, resulting in an entry-wise summarized quality of fit. 
    We find a total root-mean-square deviation of the fit of $0.14$.
    \textbf{(e)} Instantaneous root-mean-square deviation of the identified Hamiltonian $\hat h$, initial map $\hat S$ and final map $\hat D_M$ and of the target Hamiltonian $h_0$ with initial map fit $S_0$ from the experimental data averaged over the distinct time series. }
\end{figure*}

In \cref{fig:outline}, we show an overview of the experimental procedure, and the different steps of the Hamiltonian identification algorithm. 
Every experiment uses a few coupled qubits, from the larger array of qubits on the device (\cref{fig:outline}(a)).
On those qubits, the goal is to implement the time-evolution with targeted Hamiltonian parameters $h_0$, which are subject to connectivity constraints imposed by the couplings of the qubits. 
To achieve this, we perform the following pulse sequence to collect dynamical data of the form \eqref{eq:time-evolution data}.
Before the start of the sequence, the qubits are at frequencies (of the $\ket{0}$ to $\ket{1}$ transition) that could be a few hundred MHz apart from each other. 
In the beginning, all qubits are in their ground state $\ket{0}$. 
To prepare the initial state, a $\pi/2$-pulse is applied to one of the qubits, resulting in its Bloch vector moving to the equator. 
Then ramping pulses are applied to all qubits to bring them to the desired detuning around a common rendezvous frequency ($6500\,\text{MHz}$ in this work). At the same time, pulses are applied to the couplers to set the nearest-neighbour hopping to the desired value ($20\,\text{MHz}$ in this work). 
The pulses are held at the target values for time $t$, corresponding to the evolution time of the experiment. 
Subsequently, the couplers are ramped back to zero coupling and the qubits back to their initial frequency, where $ \langle x_m(t) \rangle$ and $\langle p_m(t)\rangle$ on the desired qubit $m$ is measured. 
The initial and final pulse ramping take place over a finite time of $2$--$3$ ns, and therefore give rise to a non-trivial effect on the dynamics, which we take into account in the identification procedure. 
In fact, we find that the effects of the ramping phase are the domininant source of SPAM errors in the quench-based analog simulation.
The experimental data (\cref{fig:outline}(b)) on $N$ qubits are $N \times N$ time-series estimates of $\langle a_m(t) \rangle_{\psi_n}$ for $t = 0, 1, \ldots, T$ ns and all pairs $n,m = 1, \ldots, N$. 
Given those data, the identification task amounts to identifying the `best' coefficient matrix $h$, describing the time-sequence of snapshots of the single-particle unitary matrix $\frac12 \exp(- \i t h)$. \smallskip

\paragraph*{Identification method.}
We can identify the generator $h$ of the unitary in two steps (\cref{fig:outline}(c)), making use of the eigendecomposition of the Hamiltonian (see Methods). In the first step, the time-dependent part of the identification problem is solved, namely, identifying the Hamiltonian eigenvalues (eigenfrequencies).
In the second step, given the eigenvalues, the eigenbasis for the Hamiltonian of $h$ is determined. 
In order to make the identification method 
noise-robust, we furthermore exploit 
structural constraints of the model. 
First, the Hamiltonian has a spectrum such that the time-series data has a time-independent, sparse frequency spectrum with exactly $N$ contributions.
Second, the Fourier coefficients of the data have an explicit form as the outer product of the orthogonal eigenvectors of the Hamiltonian.
Third, the Hamiltonian parameter matrix is real and has an a priori known sparse support due to the experimental connectivity constraints. 
These structural constraints are not respected by various sources of incoherent noise, including particle loss and finite shot noise, and coherent noise, in particular the SPAM error.
Thereby, an identification protocol that takes these constraints into account is intrinsically robust against various imperfections.
Importantly, we do not assume that the dynamics of the device is completely governed by a non-interacting, particle-number preserving Hamiltonian of the form \eqref{eq:non-interacting h}.
We rather impose this as a constraint on the reconstructed Hamiltonian and, as such, identify the best-fit Hamiltonian satisfying the constraint.
Our approach thus robustly identifies the non-interacting part of a potentially interacting system.

To robustly identify the sparse frequencies from the experimental data, we develop a new super-resolution and denoising algorithm tensorESPRIT that is applicable to matrix-valued time series and uses tensor network techniques in conjunction with super-resolution techniques for scalar data~\cite{Fannjiang:2016:ESPRIT}.
Achieving high precision in this step is crucial for identifying the eigenvectors in the presence of noise. To robustly identify the eigenbasis, in the second step, we perform least-square optimization of the time-series data under the orthonormality constraint with a gradient descent algorithm on the manifold structure of the orthogonal group~\cite{abrudan_conjugate_2009}. 
Additionally, we can incorporate the connectivity constraint on the coefficient matrix $h$ by making use of regularization techniques~\cite{Regularization}.

\paragraph*{Robustness against ramp errors.}
The initial and final ramping pulses result in a time-independent, linear transformation at the beginning and end of the time series.
It is important to stress that 
such ramping pulses are expected to be generic
in a wide range of experimental implementations of dynamical analog quantum simulations. 
Robustness of a Hamiltonian identification method against these imperfections is essential for 
accurate estimates in practice.
We can model the effect of such state \emph{preparation and measurement} (SPAM) errors via linear maps $S$ and $M$, which alters our model of the ideal data \eqref{eq:time-evolution data} to 
\begin{align}
\label{eq:spam time-evolution data}
\langle a_m(t) \rangle_{\psi_n}  = \frac 12(M\cdot \exp(-\i t h) \cdot S)_{m,n}.
\end{align}
These linear maps capture the effect of particle-number preserving quenches, as well as the projection of more general channels to the single-particle subspace. 
Any deviation of the observed experimental dynamics from our model of the data~\eqref{eq:spam time-evolution data} will be visible in the quality of fit.

While for the frequency identification such time-independent errors `only' deteriorate the signal-to-noise ratio, for the identification of the eigenvectors of $h$ it is crucial to take the effects of non-trivial $S$ and $M$ into account. 
Given the details of the ramping procedure, we expect that the deviation of the initial map $S$ from the identity will be significantly larger than that of the final map $M$ and provide evidence for this in the Methods. 
In particular, the final map will be dominated by phase accumulation on the diagonal. 

By pre-processing the data, we can robustly remove an arbitrary initial map $S$. 
By post-processing, we can obtain an orthogonal diagonal estimate $\hat D_M$ of the final map $M$.
We give numerical evidence that the estimate $\hat D_M$ gives good results in the particular experimental setting.
From the identified Hamiltonian and an orthogonal diagonal estimate $\hat D_M$ of $M$, we get an estimate $\hat S$ of $S$. 
\smallskip

\begin{figure}[ht!]
  \includegraphics{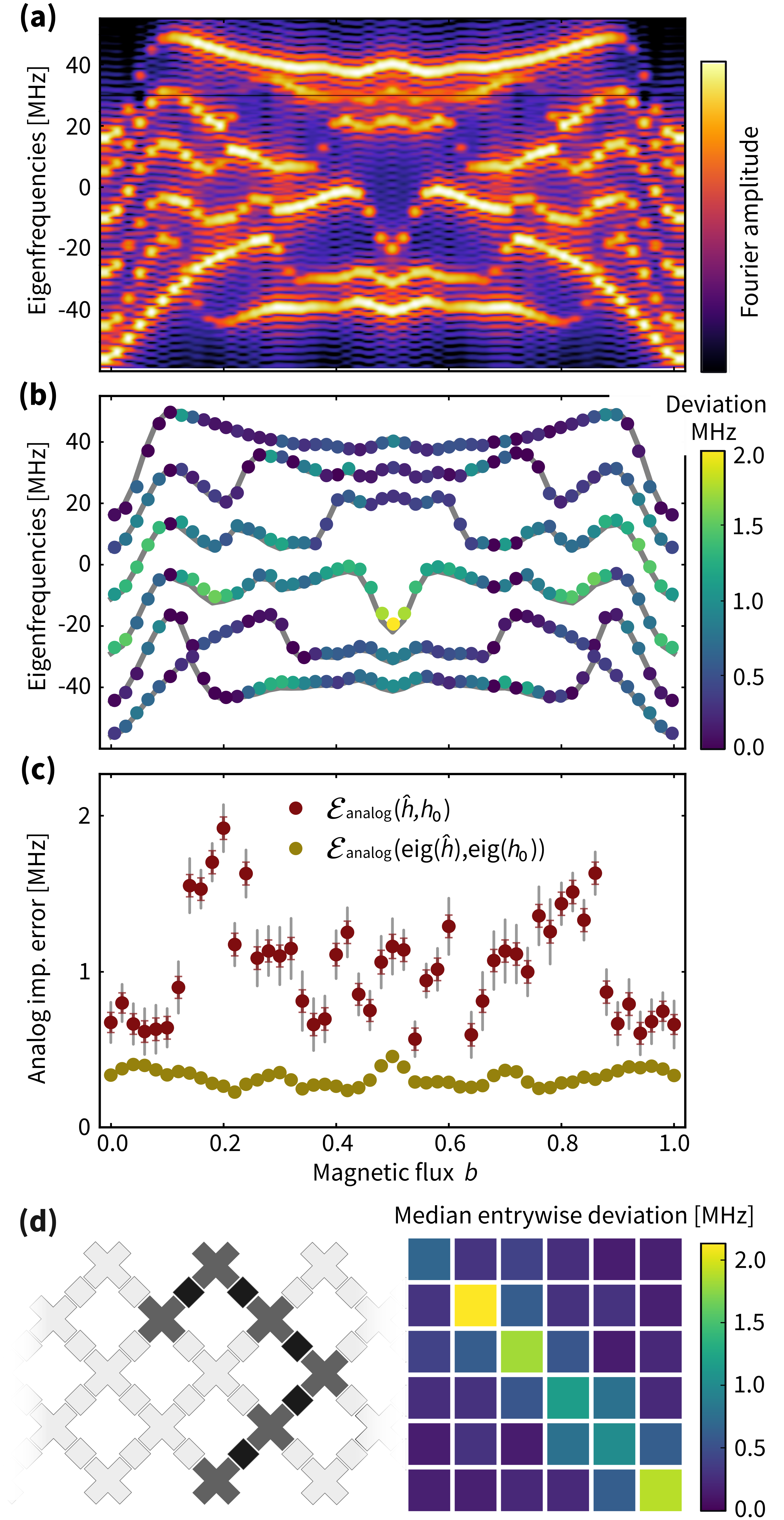}
  \caption{
    \label{fig:many H} 
    \textbf{Comparing frequency and full identification errors}. 
    \textbf{(a)} In an $N = 6$ subset of connected qubits, by varying $b$ from~$0$ to~$1$, we implement~51 different Hamiltonians. The plot shows the Fourier transform of the time domain data. 
    \textbf{(b)} The extracted eigenfrequencies (denoised peaks in panel (a)) are shown as colored dots, where the assigned color is indicative of the deviation between targeted eigenfrequencies (gray lines) and the identified ones from position of the peaks. 
    \textbf{(c)} Analog implementation error $\mc E_{\rm analog}(\hat h, h_0)$ of the identified Hamiltonian\,(dark red) compared to the implementation error $\mc E_{\rm analog}(\eig(\hat h),\eig(h_0))$ of the identified frequencies\,(golden). 
    Colored (gray) error bars quantify the statistical (systematic) error. 
    \textbf{(d)} Layout of the six qubits on the Sycamore processor and median of the entry-wise absolute-value deviation of the Hamiltonian matrix entries from their targeted values across the ensemble of $51$ different values of $b \in [0,1]$. 
     }
\end{figure}

\paragraph*{Error sources.}
There are two main remaining sources of error that affect the Hamiltonian identification.
First, the estimate $\hat h$ has a statistical error due to the finite number of measurements used to estimate the expectation values. 
Second, any non-trivial final map $M$ will produce a systematic error in the eigenbasis reconstruction and the tomographic estimate $\hat S$.
We partially remedy this effect with an orthogonal diagonal estimate $\hat D_M$ of $M$. 

\paragraph*{Predictive power.}
If the dynamics of the device is indeed coherent and particle-number preserving, the learned model will allow us to accurately predict the dynamics of the device in the single-particle subspace. 
If, additionally, interactions are negligible, the predictive power of our model extends to dynamics of more particles. 
This allows us to benchmark the Sycamore chip as a programmable quantum simulator of the non-interacting part of a Bose-Hubbard model.
Accurately predicting the dynamics of many particles requires a generalization of our method to at least the two-particle sector. 
\smallskip

\paragraph*{Results.}
We implement and characterize different Hamiltonians from time-series data on two distinct quantum Sycamore processors---\emph{Sycamore \#1 and  \#2}.
The Hamiltonians we implement have a fixed overall hopping strength $J_{i,j}=20$\,MHz and site-dependent local potentials $\mu_i$ on subsets of qubits. 
Specifically, we choose the local potentials quasi-randomly $\mu_q = 20\cos(2\pi q b)\,\text{MHz}$, for $q = 1, \ldots, N$, where $b$ is a number between zero and one. 
In one dimension, this choice corresponds to implementing the Harper Hamiltonian, which exhibits characteristic `Hofstadter butterfly' frequency spectra as a function of 
the dimensionless magnetic flux $b$~\cite{Hofstadter1976}. 
 
We measure deviations in the identification in terms of the \emph{analog implementation error} of the identified Hamiltonian $\hat h$ with respect to the targeted Hamiltonian $h_0$ as 
\begin{equation}
\label{eq:implementation error} 
\mathcal{E}_{\rm analog}(\hat h, h_0) 
\coloneqq \frac 1 N \norm{\hat{h}  - h_0}_{\ell_2}, 
\end{equation}
defined in terms of the $\ell_2$-norm,
which for a matrix $A$ is given by $\norm{A}_{\ell_2} = (\sum_{i,j} |A_{i,j}|^2)^{1/2}$. 
We also use the analog implementation error to quantify the implementation error of the initial map $\hat S$ as $\mc E_{\rm analog}(\hat S,\id)$, and of the eigenfrequencies $\eig(\hat h)$ as $\mc E_{\rm analog}(\eig (\hat h),\eig (h_0))$. 
Notice that the analog implementation error of the frequencies in the data from the targeted Hamiltonian eigenfrequencies give a lower bound to the overall implementation error of the identified Hamiltonian. 
This is because the $\ell_2$-norm used in the definition \eqref{eq:implementation error} of $\mc E_{\rm analog}$ is unitarily invariant and any deviation in the eigenbasis, which we identify in the second step of our algorithm, will tend to add up with the frequency deviation. 

In \cref{fig:single H}, we illustrate the properties of a single Hamiltonian identification instance in terms of both how well the simulated time evolution fits the experimental data (a,d,e) and how it compares to the targeted Hamiltonian~(b) and SPAM~(c).
We find that most entries of the identified Hamiltonian deviate from the target Hamiltonian by less than $0.5$ MHz with a few entries deviating by around $1$--$2$ MHz. 
The overall implementation error is around $1$ MHz. 
The error of the identification method is dominated by the systematic error due to the final ramping phase that is around $1$ MHz for the individual entries, see the SM for details. 
Small long-range couplings exceeding the statistical error are necessary to fit the data well even when penalizing those entries via regularization.  
These entries are rooted in the effective rotation by the final ramping before the measurement and within the estimated systematic error. 

The fit deviation from the data (\cref{fig:single H}(e)) exhibits a prominent decrease within the first few nanoseconds of the time evolution. 
This indicates that the time evolution differs during the initial phase of the experiment as compared to the main phase of the experiment, which we can attribute to the initial pulse ramping of the experiment.
The identified initial map describing this ramping (\cref{fig:single H}(c)) is approximately band-diagonal and deviates from being unitary, indicating fluctuations of the effective ramps between different experiments. 

We find a larger time-averaged real-time error (\cref{fig:single H}(d)) in all data series $\langle a_m \rangle_{\psi_n}$ in which $Q_4$ was measured, indicating a measurement error on $Q_4$. 
We also observe a deviation between the parameters of the target and identified Hamiltonian in qubits $Q_3$ and $Q_4$ and the coupler between them. 
Since the deviation of the eigenfrequencies is much smaller than of the full Hamiltonian, we attribute those errors also to a 
non-trivial final ramping phase at those qubits that leads to a rotated eigenbasis.

\begin{figure}[tb]
\includegraphics{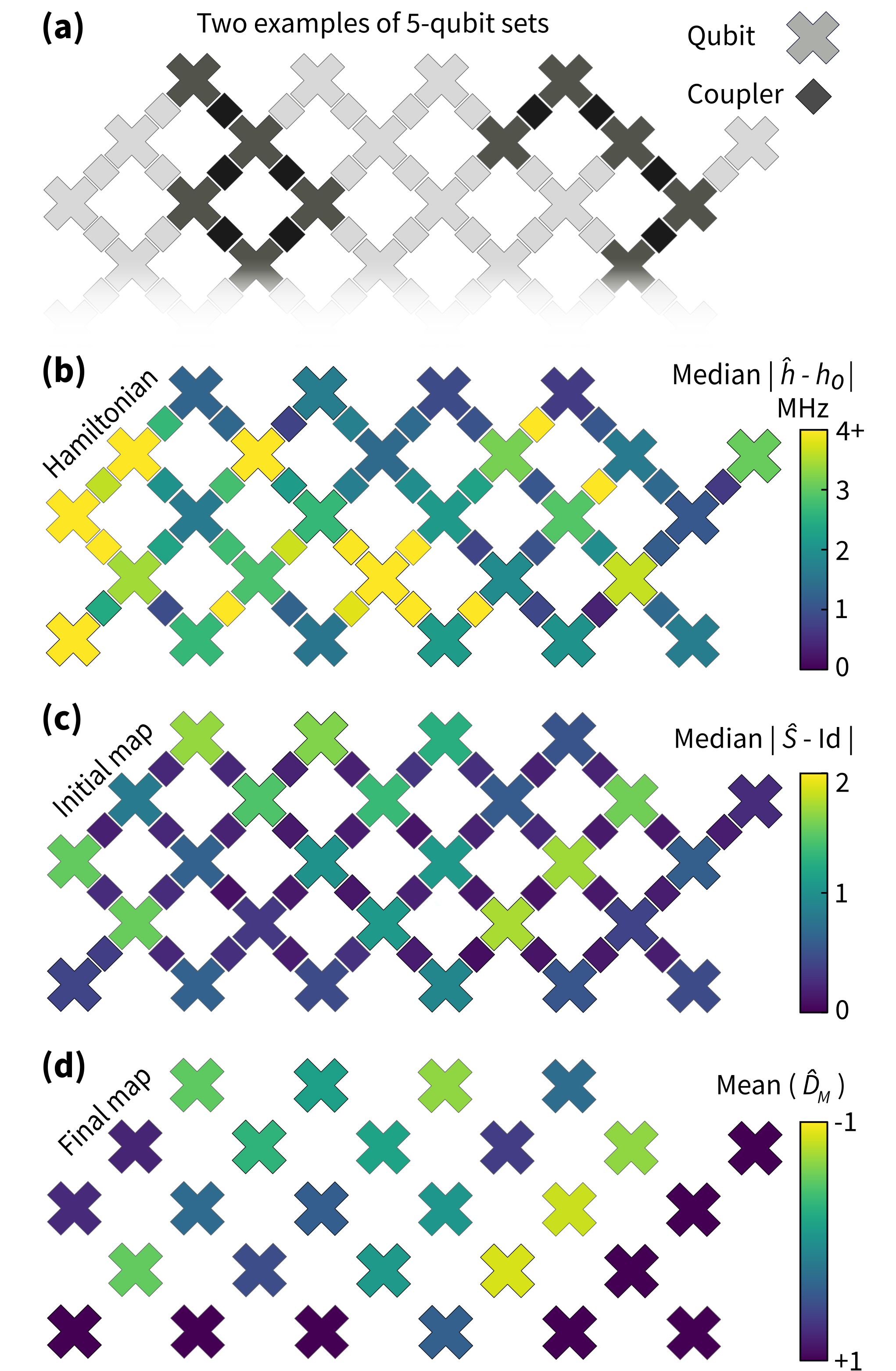}
  \caption{
    \label{fig:scan chip}
    \textbf{Error map of Hamiltonian implementation across the Sycamore processor.} Over the grid of $27$ qubits, we randomly choose subsets of connected qubits and couplers of size $N = 5$. On each subset we implement two Hamiltonians with $b = 0, 0.5$ and run the identification algorithm. Two instances are shown in panel \textbf{(a)}.  
    For each subset, we compute the deviation of the identified Hamiltonian and initial map from their respective target and assign it to each qubit or coupler involved. 
    Due to overlap of subsets, each qubit or coupler has been involved in at least $5$ different choices of subsets.
    Panels \textbf{(b)} and \textbf{(c)} show the median deviation for the Hamiltonian and initial map implementations, respectively. 
    Panel \textbf{(d)} shows the mean of the sign flips in the identified (diagonal $\pm 1$) final map for each qubit. 
       }
\end{figure}

In \cref{fig:many H}, we summarize multiple identification data of this type to benchmark the overall performance of a fixed set of qubits. 
In panel (a), we show the measured Fourier domain data for $51$ different values of the magnetic flux $b \in [0,1]$. 
In panel (b), we plot the deviation of the frequencies identified from the data. 
Most implemented frequencies deviate by less than 1\,MHz from their targets. 
Importantly, the frequency identification is robust against systematic measurement errors. 
When comparing the analog implementation errors of the full Hamiltonian (\cref{fig:many H}(c)) to the corresponding frequency errors, we find an up to fourfold increase in implementation error.
The Hamiltonian implementation error is affected by a systematic error due to the non-trivial final ramp.
We estimate this error using a linear ramping model; see the SM for details.
Since the deviation lies outside of the combined systematic and statistical error bars, our results indicate that the targeted Hamiltonian has not been implemented exactly.

In \cref{fig:many H}(d), we show 
the median of the entry-wise deviation of the identified Hamiltonian from its target over all magnetic flux values. 
Thereby, the ensemble of Hamiltonians defines an overall error benchmark. 
This benchmark can be associated to the individual constituents of the quantum processor, namely, the qubits, corresponding to diagonal entries of the Hamiltonian deviation, and the couplers, corresponding to the first off-diagonal matrix entries of the deviation. 

We use this benchmark over an ensemble of two flux values to assess a $27$-qubit array of superconducting qubits. To do so, we repeat the analysis reported in \cref{fig:many H} for $5$-qubit dynamics on different subsets of qubits and extract average errors of the individual qubits and couplers involved in the dynamics, both in terms of the identified Hamiltonian and the initial and final maps. 
Summarized in \cref{fig:scan chip}, we find significant variation in the implementation error of different couplers and qubits. 
While for some qubits the effects of the initial and final maps are negligible, for others they indicate the potential of a significant implementation error.
From a practical point of view, such diagnostic data allows to maximally exploit the chip's error for small-scale analog simulation experiments.
Let us note that within the error of our method the overall benchmark for the qubits and couplers for $5$-qubit dynamics agrees with that of $3$- and $4$-qubit dynamics. 

\begin{figure}
    \includegraphics{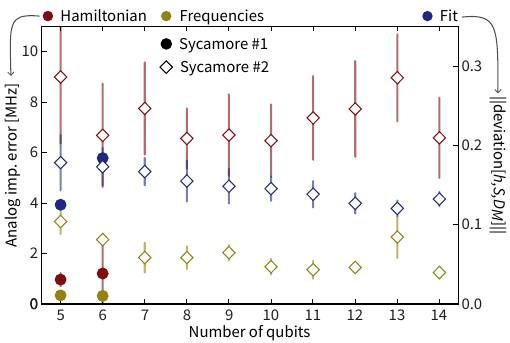}
    \caption{\label{fig:increasing system size}
    \textbf{Analog implementation error scaling and comparing different quantum processors. }
    We measure the analog implementation error 
    of the implemented Hamiltonians (dark red) and their eigenfrequencies 
    (golden) as well as the deviation $(\sum_{l=0}^L \| \textsf{deviation}[\hat h, \hat S, \hat D_M](t_l) \|_{\ell_2}^2/(N^2 (L+1)))^{1/2}$ of the fit from the experimental data (dark blue) all averaged over implementations of Hamiltonians with quasi-random local potential on an increasing number of qubits on two different quantum processors---\emph{Sycamore \#1} (circles) and \emph{\#2} (diamonds). 
    Each point is the mean of the respective quantity over 51 Hamiltonian implementations (21 for $N=5$ and 20 for $N=14$ on Sycamore \#2). 
    The data points at $N=6$ on Sycamore \#1 summarizes \cref{fig:many H}(c).
    The error bars represent one standard deviation.
    }
\end{figure}

All of the Hamiltonian identification experiments discussed so far (Figs.~\ref{fig:single H}, \ref{fig:many H}, \ref{fig:scan chip}) were implemented on the Sycamore \#1 chip. 
In order to compare these results to implementations on a physically distinct chip with different calibration, and to demonstrate the scalability of our method, we implement Hamiltonian identification experiments for an increasing number of qubits on the Sycamore \#2 chip. 
More precisely, for a given number of qubits $N$, we implement many different Hamiltonians with quasi-random local potentials, as shown in \cref{fig:many H}(c) for $N=6$. 
We then average the analog implementation errors of the Hamiltonians and frequencies for several system sizes.  
The results are shown in \cref{fig:increasing system size}. 
Notably, comparing the two different processors, the overall quality of fit does not depend significantly on either the number of qubits or the processor used. 
This indicates, first, that our reconstruction method works equally well in all scenarios and, second, that both quantum processors implement Hamiltonian time evolution that closely fits our model assumption. 
We also notice that the overall analog implementation error does not significantly depend on the system size.
This signifies that no additional non-local errors are introduced into the system 
as the size is increased. 
At the same time, the overall error of Hamiltonian implementations on Sycamore \#2 is much worse compared to those on Sycamore \#1, indicating that Sycamore \#2 was not as well calibrated.
Hamiltonian identification thus allows us to meaningfully compare Hamiltonian 
implementations across different physical systems and system sizes.
\smallskip

\paragraph*{Conclusion.}
We have implemented analog simulation of the time-evolution of non-interacting bosonic Hamiltonians with tunable parameters for up to $14$ qubit lattice sites. 
A structure-exploiting learning method allows us to robustly identify the implemented Hamiltonian that governs the time-evolution. 
To achieve this, we have introduced a new super-resolution algorithm, referred to as tensorESPRIT, for precise robust identification of eigenfrequencies of a Hermitian matrix from noisy snapshots of the one parameter unitary subgroup it generates. 
Thereby, we diagnose the deviation from the target Hamiltonian and assess the precision of the implementation. 
We achieve sub-MHz error of the Hamiltonian parameters compared to their targeted values in most implementations. 
Combining the average performance measures over ensembles of Hamiltonians we associate benchmarks to the components of the superconducting qubit chips that quantify the performance of the hardware on the time evolution and provide specific diagnostic information.
Within our Hamiltonian identification framework, we are able to identify SPAM errors due to parameter ramp phases as a severe limitation of the architecture.
Importantly, such ramp phases are present in \emph{any} analog quantum simulation of quenched dynamics. 
Our results show that minimizing those is crucial for precisely implementing a Hamiltonian. 

The experimental and computational effort of the identification method scales efficiently in the number of modes of the Hamiltonian. 
We have also numerically identified the limitations of more direct algorithmic approaches and demonstrated the scalability of our method under empirically derived noise and error models.

We have demonstrated and custom-tailored our approach here to a superconducting analog quantum simulation platform. 
It can be applied directly to any bosonic and fermionic analog simulation platform which allows for accurate preparation and measurement of single particle excitations at specific lattice sites. 
Generalizing our two-step approach developed here, we expect a polynomial scaling with the dimension of the diagnosed particle sector and therefore remain efficient for diagnosing two-, three- and four-body interactions, thus allowing to build trust in the correct implementation of interacting Hamiltonian dynamics as a whole. 
Furthermore, it is in some cases possible to adapt the method to Hamiltonians with general non-particle number preserving free part.
From a broader perspective, with this work, we hope to contribute to the development of a machinery for precisely characterizing and thereby improving analog quantum devices. 

\let\oldaddcontentsline\addcontentsline
\renewcommand{\addcontentsline}[3]{}
\section*{Methods}

\subsection{Experimental details}

\paragraph*{Details on the quantum processor.}
We use the Sycamore quantum processor composed of quantum systems
arranged in a two-dimensional array. This processor consists of gmon qubits (transmons with tunable coupling) with frequencies ranging
from $5$ to $7$ GHz. These frequencies are chosen to mitigate a
variety of error mechanisms such as two-level defects. Our coupler design allows us to quickly tune the qubit-qubit coupling from $0$ to $40+$ MHz. The chip is connected to a superconducting circuit board and cooled down to below $20$ mK in a dilution refrigerator. 
The median values of the $T_1$ and $T_2$ times of the qubits are $T_1 = 16.1$ \textmu s, $T_2 = 5.3$ \textmu s (Ramsey interferometry) and $T_2 = 17.8$ \textmu s (after CPMG dynamical decoupling).
Each qubit has a microwave control line used to drive an excitation and a flux control line to tune the frequency. The processor is connected through filters to room-temperature electronics that synthesize the control signals. We execute single-qubit gates by driving 25 ns microwave pulses resonant with the qubit transition frequency, resulting in single-qubit gate fieldity of 99.8\% as measured via randomized benchmarking. \smallskip

\paragraph*{Ramping pulses.}
The pulses used in the experiment are pre-distorted in order to compensate for filters on the control lines. In order to calibrate this distortion, we send rectangular pulses to the qubits and monitor the frequency change of the qubits. This allows us to know the response of the microwave lines at the qubits (i.e., the deviation from a rectangle) and compensate for distortions. The ramp time can be as fast as 2 to 3 ns and the distortions take the form of overshoot and undershoots with a long response time of ~100ns. After compensating for the distortions, the qubit frequency remains fixed. 
\smallskip

\paragraph*{Experimental read-out and control.} The qubits are connected to a resonator that is used to read out the state of the qubit. The state of all qubits can be read simultaneously by using a frequency-multiplexing. Initial device calibration is performed using `Optimus' \cite{kelly_physical_2018} where calibration experiments are represented as nodes in a graph.

\subsection{Details of the identification algorithm}
Succinctly written, our data model is
given by 
\begin{align}
\label{eq:data model methods}
	y_{m,n}[l] \coloneqq \langle a_m(t_l)\rangle_{\psi_n} = \frac 12 (M \cdot \exp(- \i t_l h) \cdot S)_{m,n}, 
\end{align}
where $m,n = 1, \ldots, N$ label the distinct time series, 
$l = 0, \ldots, L$ 
labels the time stamps of the $L+1$ data points per time series. 
The matrices $S$ and $M$ are arbitrary invertible linear maps that capture the state preparation and measurement stage, as affected by the ramping of the eigenfrequencies of the qubits and couplers to their target value and back (see \cref{fig:outline}).
In the experiment, we empirically estimate each such expectation value with 1000 single shots.

Our mindset for solving the identification problem is based on the eigendecomposition $h = \sum_{k= 1}^N \lambda_k \proj {v_k}$ of the coefficient matrix $h$ in terms of eigenvectors $\ket{v_k}$ and eigenvalues $\lambda_k$. 
We can write the data \eqref{eq:data model methods} in matrix form as 
\begin{align}
  \label{eq:diagonalized data}
  y[l] & = \frac 12 \exp(-\i t_l h) 
  = \frac 12 \sum_{k = 1}^N \e^{-\i t_l \lambda_k} \proj{v_k}\, , 
\end{align}
where we have dropped $S$ and $M$ for the time being.
This decomposition suggests a simple procedure to identify the Hamiltonian using Fourier data analysis. 
From the matrix-valued time series data $y[l]$ \eqref{eq:diagonalized data}, we identify the Hamiltonian coefficient matrix $h$ in two steps. 
First, we determine the eigenfrequencies of $h$. 
Second, we identify the eigenbasis of $h$.
To achieve those identification tasks with the largest possible robustness to error, it is key to exploit all available structure at hand. \smallskip

\paragraph*{Step 1: Frequency extraction.} 
In order to robustly estimate the spectrum, we exploit that the signal is \emph{sparse} in Fourier space. 
This structure allows us to substantially denoise the signal and achieve \emph{super-resolution} beyond the Nyquist limit \cite{CandesFernandez-Granda:2013, CandesFernandez-Granda:2014}. 
A candidate algorithm for this task, suitable for scalar time-series, is the ESPRIT algorithm, which comes with rigorous recovery guarantees~\cite{Fannjiang:2016:ESPRIT,LiLiaoFannjiang:2019:Super-resolution}.
To extract the Hamiltonian spectrum from the matrix time-series $y[l]$, we apply ESPRIT  to the trace of the data series (for $S = M = \id$)
\begin{align}\label{eq:Sseries}
   F[l] \coloneqq &\tr[ y [l]] = \sum_{m = 1}^N y_{m,m}[l] = \frac12 \sum_{k=1}^N \e^{-\i t_l \lambda_k}\,.
\end{align}
The drawback of this approach is that if the spectrum of the Hamiltonian is sufficiently crowded, which will happen for large $N$, the Fourier modes in $F[l]$ become indistinguishable and ESPRIT fails to identify the frequencies.
In particular, ESPRIT is not able to identify degeneracies in the spectrum.

To overcome this issue and obtain a truly scalable learning procedure applicable to degenerate spectra, we develop a new algorithm coined tensorESPRIT.
TensorESPRIT extends the ideas of ESPRIT to the case of matrix-valued time series using tensor network techniques.
Using tensorESPRIT also improves the robustness of frequency estimation to SPAM errors.
For practical Hamiltonians, tensorESPRIT becomes necessary for systems with $N \gtrsim 12$; as we find in numerical simulations summarized in \cref{sec:benchmarking methods} and detail in Section IV.B of the SM.

TensorESPRIT (ESPRIT) comprises of a denoising step, in which the rank of the Hankel tensor (matrix) of the data is limited to its theoretical value.
Subsequently, rotational invariance of the data is used to compute a matrix from the denoised Hankel tensor (matrix), the spectrum of which has a simple relation to the spectrum of $h$.
In the case of ESPRIT, this amounts to a multiplication of the denoised Hankel matrix by a pseudoinverse of its shifted version.
Contrastingly, tensorESPRIT uses a sampling procedure to contract certain sub-matrices of the denoised Hankel tensor with the pseudoinverse of other sub-matrices.
Details on both algorithms can be found in the SM.  \smallskip



\paragraph*{Step 2: Eigenspace identification.}
To identify the eigenspaces of the Hamiltonian, we use the eigenfrequencies found in Step~1 to fix the oscillating part of the dynamics in \cref{eq:diagonalized data}. 
What remains is the problem of finding the eigenspaces $\proj{v_k}$ from the data. 
This problem is a non-convex inverse quadratic problem, subject to orthogonality of the eigenspaces, as well as the constraint that the resulting Hamiltonian matrix respects the connectivity of the superconducting architecture. 
Formally, we denote the a priori known support set of the Hamiltonian matrix as $\Omega$, so that we can write the support constraint as $h_{\bar{\Omega}} = 0 $, where $\bar{\Omega}$ denotes the complement of $\Omega$ and subscripting a matrix with a support set restricts the matrix to this set. 
We can cast this problem into the form of a  least-squares optimization problem
\begin{equation}\label{eq:generalEigReconstruction}
\begin{split}
\operatorname*{minimise}_{\{\ket{v_k}\}}&\quad \sum_{l=0}^L \left\| y[l] - \sum_k \e^{-\i \lambda_k t_l} \ketbra{v_k}{v_k}\right\|^2_{\ell_2} , \\
\operatorname{subject\ to}&\quad \braket{v_m}{v_n}= \delta_{m,n}, \, \left(\sum_k \lambda_k \ketbra{v_k}{v_k}\right)_{\bar\Omega} = 0,
\end{split}\end{equation}
equipped with non-convex constraints enforcing orthogonality, and the quadratic constraint restricting the support. 
In order to approximately enforce the support constraint, we make use of regularization~\cite{Regularization}.
It turns out that this can be best achieved by adding a term~\cite[App.~A]{HangleiterEtAl:2020}
\begin{align}
  \label{eq:regularization term}
\mu \norm{ \left(\sum_k \lambda_k \proj{v_k} \right)_{\bar{\Omega}}}_{\ell_2}
\end{align}
to the objective function \eqref{eq:generalEigReconstruction}, where $\mu > 0$ is a parameter weighting the violation of the support constraint. 
We then solve the resulting minimization problem by using a conjugate gradient descent on the manifold of the orthogonal group \cite{EdelmanAriasSmith:1998, abrudan_conjugate_2009}, see also the recent work \cite{luchnikov2020qgopt,luchnikov2020riemannian, RothEtAl:2020} for the use of geometric optimization for quantum characterization. 

Without the support constraint this gives rise to an optimization algorithm that converges well, as shown in the SM.
However, the regularization term makes the optimization landscape rugged as it introduces an entry-wise constraint that is skew to the orthogonal manifold. 
To deal with this, we consecutively ramp up $\mu$ until the algorithm does not converge anymore in order to find the Hamiltonian that best approximates the support constraint while being a proper solution of the optimization problem. 
For example, for the data in \cref{fig:single H} the value of $\mu$ is $121$.
In order to avoid that we identify a Hamiltonian from a local minimum of the rugged landscape, we only accept Hamiltonians that achieve a total fit of the experimental data within a 5\% margin of the fit quality of the unregularized recovery problem, and use the Hamiltonian recovered without the regularization otherwise.

\subsection{Robustness to state preparation and measurement errors}\label{ssec:spam robustness}

The experimental design requires a ramping phase of the qubit and coupler frequencies from their idle location to the desired target Hamiltonian and back for the measurement. 
In effect, the data model \eqref{eq:data model methods} includes time-independent linear maps $M$ and $S$ that are applied at the beginning and end of the Hamiltonian time-evolution.  
The maps affect both the frequency extraction and the eigenspace reconstruction. 

For the frequency extraction using ESPRIT, the Fourier coefficients of the trace signal $F[l]$ become $\sandwich{v_k}{SM}{v_k}$.
While the frequencies remain unchanged the Fourier coefficients now deviate from unity, significantly impairing the noise-robustness of the frequency identification.  
This effect is still present, albeit weaker, in tensorESPRIT, in the case of non-unitary SPAM errors.
The eigenspace reconstruction is affected much more severely and requires careful consideration, as detailed below and in the SM.

\paragraph*{Ramp removal via pre-processing.}
We can remove either the initial map $S$ or the final map $M$ from the data.
To remove $S$, we apply the pseudoinverse $(\cdot )^+$ of the data $y[l_0]$ at a fixed time $t_{l_0}$ to the entire (time-dependent) data series in matrix form. 
For invertible $S$ and $M$ this gives rise to 
\begin{equation}
  \label{eq:pre-processing ramp removal}
  \begin{split}
    y^{(l_0)}[l] &= y[l](y[l_0])^+ \\
    &= \sum_{k=1}^N \e^{-\i \lambda_k (t_l - t_{l_0})} M \ketbra{v_k}{v_k} M^{-1}. 
  \end{split}
\end{equation}
The caveat of this approach is that the shot noise that affected the single time point $y[l_0]$ can lead to correlated errors in every entry of the new data series $y^{(l_0)}$.

We can reduce the error induced by these correlations by effectively averaging over `corrected' data series $y^{(l_0)}$ with different $l_0$.  
To this end, we compute the concatenation of data series for different choices of $l_0$, e.g., for every $s$ data points $0,s,2s, \ldots, \lfloor L/s \rfloor s $ giving rise to new data $y_\text{total, s} = (y^{(0)} , y^{(s)}, y^{(2s)}, \ldots, y^{(\lfloor L/ s \rfloor s)}) \in \CC^{\lfloor L/ s \rfloor L}$.  
If the data suffers from drift errors, it is also beneficial to 
restrict each data series $y^{(l_0)}$ to entries $y^{(l_0)}[\kappa]$ with $\kappa \in [l_0 - w, l_0 + w]$, i.e., the entries in a window of size $w$ around $l_0$. 
In practice, we use $s = 1$ and $w = 50$ for the reconstructions on Sycamore \#1, and $s=1, w=L$ for 
those
on Sycamore \#2. 

As we argue below, the final map $M$ is nearly diagonal here.
Hence, we can use $y_{\text{total, s}}$ from 
Eq.~\eqref{eq:generalEigReconstruction} and it is justified to apply the support constraint in the eigenspace reconstruction step.
However, the eigenspace reconstruction will suffer from systematic errors due to the final map, even in the case when it is nearly diagonal.
Below, we explain a method to partially remove this error.
\smallskip 

\begin{figure}
  \centering
  \includegraphics{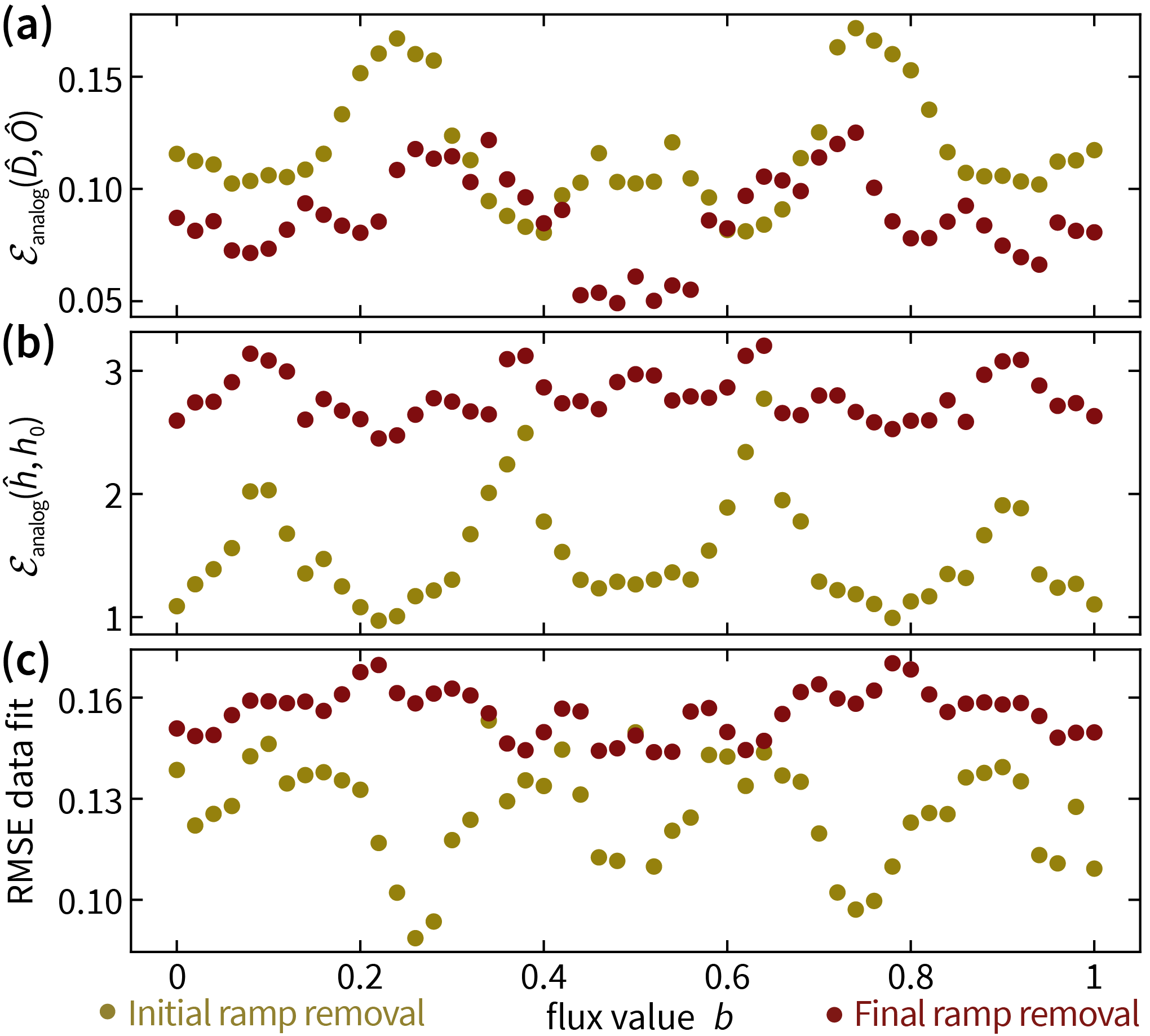}
  \caption{\label{fig:initial vs final}
  \textbf{Initial ramp removal versus final ramp removal.} We identify Hamiltonians of a set of $5$-qubit Hamiltonians with Hofstadter butterfly potentials $\mu_q = 20 \cos(2 \pi q b)$ MHz for qubits $q = 1, \ldots, 5$ and flux value $b$ in without regularization. 
  \textbf{(a)} Deviation of the orthogonal part $\hat O_S$ ($\hat O_M$) of the identified initial map $\hat S$ 
  (final map $\hat M$) from the closest diagonal orthogonal matrix $\hat D_S$ ($\hat D_M$).
  \textbf{(b)} Analog implementation error of the corresponding identified Hamiltonians $\hat h_S$ ($\hat h_M$).
  \textbf{(c)} Total root-mean-square deviation of the time series data from the Hamiltonian fit.
  }
\end{figure}
\begin{figure}
    \includegraphics{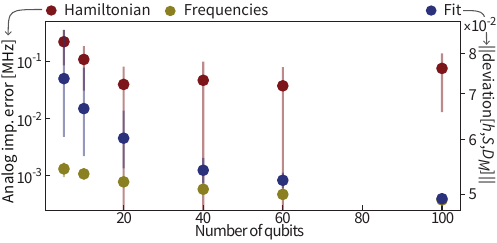}
    \caption{\label{fig:benchmarks}
    \textbf{Numerical benchmarks for larger system sizes. }
    Recovery error of frequencies (golden) and Hamiltonians (red) from simulated time series averaged over $20$ instances of Harper Hamiltonians for different system sizes. The error bars represent one standard deviation. The evolution is simulated for up to $0.6$ $\mu\text{s}$ and sampled at a rate of $250$ MHz. Statistical noise is simulated using $10^3$ shots per expected value and SPAM is modeled by using randomly chosen idle qubit and coupler frequencies, linear ramping of $1.5$ GHz/s padded by $0.05$ ns.  The fitting error of the time series is depicted in blue, right $y$-axis. 
    We refer to the SM, Sec.~VII~A for details.
    }
\end{figure}

\paragraph*{Tomographic estimate of $S$ and $M$.}
The systematic error in the reconstructed Hamiltonian eigenbasis can be expressed as an orthogonal rotation $D_M$ from the eigenbasis that is actually implemented.
Due to the gauge freedom in the model \eqref{eq:data model methods}, we cannot hope to identify $D_M$ fully without additional assumptions. 
However, as elaborated on in the SM, we can find a diagonal orthogonal estimate $\hat D_M$ of the true correction $D_M$ and hence remove a sign of the systematic error.
To this end, we assume that the experimental implementation of the target Hamiltonian does not flip the sign in the hopping terms and remedy the sign of systematic error due to the final map by fitting a diagonal orthogonal rotation of the Hamiltonian eigenbasis $\hat D_M$ that minimizes the implementation error.
We update the reconstructed Hamiltonian to 
\begin{equation}
  \hat h = \hat D_M \tilde h \hat D_M,
\end{equation}
where $\tilde h = \sum_k \lambda_k \proj{v_k}$ and $\{\ket{v_k}\}$ is the eigenbasis obtained by solving the problem \eqref{eq:generalEigReconstruction}, and use $\hat D_M$ as an estimate of $M$.
We can now obtain a tomographic estimate of the initial map through
\begin{equation}
\label{eq:identified initial map}
  \hat S \coloneqq \frac{2}{L+1}\sum_{l = 0}^L \exp[\i t_l \hat h] \hat D_M y[l]\, .
\end{equation}
The recovered model $(\hat h, \hat S, \hat D_M)$ gives good prediction accuracy on simulated data, as demonstrated in \cref{fig:benchmarks} and in the SM, and fits well the experimental data, as demonstrated in \cref{fig:single H,fig:increasing system size,fig:initial vs final}.

\paragraph*{Imbalance between initial and final ramping phase.} As explained above, the pre-processing step allow us to remove either the initial map $S$ or final map $M$ from the data, 
{while we can only find a diagonal orthogonal estimate of the remaining map.}
A priori it is unclear which one of the two maps should be removed in order to reduce the systematic error more. 

We have already treated the initial and final ramping phases on a different footing, however. 
The reason for this is rooted in the specifics of the ramping of the couplers compared to the qubits. 
The couplers need to be ramped from their idle frequencies to provide the desired target frequencies of $20$\,MHz. 
This is why we expect the time scale of the initial ramping to be mainly determined by the couplers, namely the delay until they arrive around the target frequency and the time it takes to stabilize at the target frequency. 
In contrast, the final ramping map becomes effectively diagonal as soon as the couplers are again out of the MHz regime. 
We therefore expect that the initial map has a sizeable non-diagonal orthogonal component, whereas the final map is approximately diagonal. 

We build trust in this assumption using experimental data in \cref{fig:initial vs final}.
We observe that the deviation of the orthogonal part $\hat O_S$ of the identified initial map $\hat S$ from its projection $\hat D_S$ to diagonal orthogonal matrices is much larger than the corresponding deviation for the final map (\cref{fig:initial vs final}(a)). 
Moreover, both the root-mean-square fit of the data (\cref{fig:initial vs final}(c)) and the analog implementation error of the identified Hamiltonian with its target (\cref{fig:initial vs final}(b)) are significantly improved when removing the initial ramp, as compared to removing the final ramp. 
This indicates that $S$ induces a larger systematic error than $M$. 
Correspondingly, it is indeed more advantageous to remove the initial map in the pre-processing and fit the final map with a diagonal orthogonal matrix, validating the approach taken here. 
In the SM, we provide further numerical evidence that this approach leads to small systematic errors and recovers a model with good predictive power.

\smallskip

\subsection{Benchmarking the algorithm}
\label{sec:benchmarking methods}

We benchmark our identification algorithm against more direct approaches in numerical simulations including models for statistical and systematic errors in the SM~VI. 
We find that, indeed, already for small system sizes, the regularized manifold optimization algorithm developed here features an
improved robustness against state preparation and measurement errors compared to (post-projected) linear inversion. 
For intermediate system sizes ($N > 10$), exploiting structure in the recovery algorithm then becomes an imperative. 
In particular, for larger system sizes the eigenspectrum of the Hamiltonian becomes unavoidably narrower spaced, leading to (near-)degeneracies. 
We find that on instances of the Harper Hamiltonian studied here linear inversion approaches cannot be applied at all for $N>20$. Regularized conjugate gradient decent in contrast yields good recovery performance even for larger systems. 
The same limitations apply to a direct Fourier analysis of the cumulative
time series data using ESPRIT, as described above.  For different families of Hamiltonians, we find that above a system size of $N\approx20$ tensorESPRIT still consistently recovers the frequency spectrum, while the ESPRIT algorithm fails to do so.

Using structure not only allows our algorithm to denoise the data and achieve error robustness, it also makes precise Hamiltonian identification possible even with the number of measurements dramatically reduced in the spirit of compressed sensing.
As described above, the number of measurements scales quadratically with the system size. 
We find that
using the conjugate gradient algorithm the identification procedure reliably recovers Hamiltonians even when it has access to only about 3\% of the measurements.
In this regime, the linear inverse problem of finding the eigenvectors is underdetermined. 
Thus, the required experimental resources can be significantly reduced for large system sizes.  

To demonstrate our method's scalability, \cref{fig:benchmarks} shows the recovery performance of the structure-exploiting algorithm on simulated data under realistic models for SPAM errors and with finite measurement statistics in the regime where the baseline approaches could not be applied anymore. 
 
As detailed in the SM, tensorESPRIT has computational complexity in $\mathcal O(L^2 N^3)$.  
It is not straight-forward to bound the computational complexity of the conjugate gradient descent, as it depends on the required precision of the matrix exponential and the number of descent steps until convergence.
The entire identification algorithm consumes $\mathcal O(LN^2)$ memory.
In practice, we find that the algorithm can be easily deployed on a consumer-grade laptop computer, e.g.\ reconstructing Hamiltonians of size $N=50$ in around $5$ minutes.

\subsection{Error estimation} \label{sec:error estimation}

We here discuss how we estimate the systematic and statistical contributions to the error on the identified Hamiltonian $\hat h$ and initial map $\hat S$.
Note that the impact of the systematic error on predicting results of experiments with the same initial and final ramps is reduced due to the gauge invariance of the model \eqref{eq:data model methods}.
Due to this freedom, some of the error in identifying $\hat D_M \sim M$ gets accounted for by a corresponding error in the identification $\hat h \sim h$ and $\hat S \sim S$ in expressions of the type $\hat M e^{-it\hat h} \hat S$.
This prediction error can be further decreased by running the algorithm twice---removing the initial map in the first run and the final map in the second run, using the first ramp estimates to partially remove the ramps from the data before running the second iteration of the identification.
This procedure is detailed and supported by numerical evidence in the SM.

\paragraph*{Systematic error: Final ramp effect estimation.}

In order to estimate the magnitude of the systematic error that is induced by the non-trivial final map, we use a linear model of the final ramping phase with a constant ramping speed and constant wait time between the coupler and qubit ramping.
We detail and present validation of this ramping model with a separate experiment in the SM, where we also provide empirical estimates of the model parameters.

Given a Hamiltonian matrix $\hat h$ and the initial ramp $\hat S$ obtained from experimental data, we recover the Hamiltonian matrix $\hat h'$ from data simulated using the model $(\hat h,\hat S,M)$, where $M$ is the final ramp given by our ramping model.
We use $|f(\hat h) - f(\hat h')|$ as an estimate of the systematic error on quantities of the form $f(\hat h) \in \RR$, .\smallskip

\paragraph*{Statistical error: Bootstrapping.}
We estimate the effect of finite measurement statistics on the Hamiltonian estimate that is returned by the identification method via parametric bootstrapping. To this end, we simulate time series data with statistical noise using Haar-random unitaries $S$ as initial ramps, the identified Hamiltonian $\hat h$ and final ramp $M =\id$, as detailed in the SM.\smallskip

\paragraph*{Acknowledgements.} 
We acknowledge contributions from Charles Neill, Kostyantyn Kechedzhi, and Alexander Korotkov to the calibration procedure used in this analog approach. We would like to thank Christian Krumnow, Benjamin Chiaro, Alireza Seif, Markus Heinrich, and Juani Bermejo-Vega for fruitful discussions in early stages of the project. The hardware used for this experiment was developed by the Google Quantum AI hardware team, under the direction of Anthony Megrant, Julian Kelly and Yu Chen. 

\paragraph*{Funding.} 
D.~H.\ acknowledges funding from the U.S.\ Department of Defense through a QuICS Hartree fellowship. This work has been supported by the BMBF (DAQC), for which it provides benchmarking tools for analog-digital superconducting quantum devices, as well as by the DFG (specifically EI 519 20-1 on notions of Hamiltonian learning, but also CRC 183 and {GRK 2433 Deadalus}). We have also received funding  from  the  European Union's Horizon2020 research and innovation programme  
({PASQuanS2}) on programmable quantum simulators, {and the ERC (DebuQC).} \smallskip

\paragraph*{Author contributions.}
D.H.\ and I.R.\ conceived of the Hamiltonian identification 
algorithm. J.F.\ conceived of the tensorESPRIT algorithm.
D.H., I.R.\ and J.F.\ analyzed the experimental data and benchmarked the identification algorithm. P.R.\ took the experimental data. 
D.H. and I.R.\ wrote the initial manuscript.
D.H., I.R., J.F., J.E. and P.R. contributed to discussions and writing the final manuscript.\smallskip

\paragraph*{Data and materials availability.} The experimental data is available from the authors upon reasonable request.\smallskip

\paragraph*{Conflict of interest.} The 
authors declare no conflict of interest.
\putbib
\let\addcontentsline\oldaddcontentsline%

\end{bibunit}

\onecolumngrid
\cleardoublepage
\setcounter{section}{0}
\setcounter{page}{0}
\setcounter{equation}{0}
\thispagestyle{empty}
\begin{center}

\textbf{\large Supplemental material for\\ 
``Robustly learning the Hamiltonian dynamics of a superconducting quantum processor''}\\
\vspace{2ex}

Dominik Hangleiter, Ingo Roth, Jon\'a\v{s} Fuksa, Jens Eisert, Pedram Roushan

\vspace{2ex}
\end{center}

\begin{bibunit}

\twocolumngrid

\renewcommand\thesection{S\arabic{section}}

\pagenumbering{roman}
\renewcommand\thefigure{S\arabic{figure}}
\renewcommand\theequation{S\arabic{equation}}

\tableofcontents


\section{Overview}
\label{sec:overview}

In this supplemental material, we elaborate on the details of the identification algorithm, provide numerical benchmarks for it, and discuss in more detail how we estimate the statistical and systematic errors on the Hamiltonian identified in the experiment. 

Recall that our ideal data model---in the absence of \emph{state-preparation and measurement} (SPAM) errors is given by 
(cf.\ \cref{eq:time-evolution data}  of the main text) 
\begin{align}
    \label{eq:data model}
    y[l] = \frac 1 2 \exp(- \i t_l h), 
\end{align}
where $t_l, \, l = 0, \ldots, L$ are the time stamps, and $h$ is the $N \times N$ coefficient matrix of the non-interacting Hamiltonian given in \cref{eq:non-interacting h}. 
Our identification technique follows the standard approach of Fourier analysis of time series signals but at the same time aims to maximally exploit all structure present in the signal. 
Our starting point is the eigendecomposition $h = \sum_{k= 1}^N \lambda_k \ket {v_k}\bra{v_k}$ of the coefficient matrix $h$, where $\lambda_k$ are the eigenvalues and $\ket {v_k}$ the eigenvectors. 
Using the eigendecomposition we diagonalise the signal as
\begin{align}
  \notag
      y[l] & = \frac 12 \exp(-\i t_l h)\\\label{eq:diagonalized data}
      & = \frac 12 \sum_{k = 1}^N \e^{-\i t_l \lambda_k} \proj{v_k}. 
\end{align}
From a signal processing perspective, the time series of each entry $y_{m,n}$ is therefore  given by a complex linear superposition of a small number of sinusoids. 
But additional constraints relate the different time series. 
As discussed in the main text, our identification technique proceeds in two steps:
In the \emph{first step}, we estimate the frequency spectrum of the time series, i.e., the eigenvalues $\lambda_k$ of $h$.  
Crucially, we can exploit that the signal is sparse in Fourier space and also that $y[l]$ are noisy samples from a one-parameter subgroup of the unitary group to dramatically denoise the signal and arrive at sub-Nyquist resolution.  
Given the frequency spectrum of the time series data, the estimation of the Fourier coefficients becomes a linear inverse problem.
In standard Fourier data analysis the Fourier coefficients are therefore typically inferred via linear inversion in the \emph{second step}. 
However, in the case of Hamiltonian recovery, the problem has a considerably richer structure that we can take advantage of.
The $N^3$ Fourier coefficients are certain second order polynomials of the entries of a set of \emph{orthogonal vectors} $\{\ket{v_k}\}^N_{k=1}$.  
Furthermore, if the interaction graph is not fully connected, the resulting sparsity pattern of the Hamiltonian matrix gives rise to linear constraints on the coefficients. 

As we will see, exploiting the structure in the second step brings two benefits in terms of noise robustness:  
First, explicitly enforcing the orthogonality constraint, polynomial structure (low-rank) and the linear constraints from the interaction graph renders the reconstruction significantly more robust against different sources of errors such as incoherent measurement errors. 
Second, since the coefficient matrices for each frequency are unit rank projectors, we can significantly denoise the signal from systematic (state-preparation) errors on the projectors by restricting ourselves to inferring the projector's range first, i.e., $\{\ket{v_k}\}_{k=1}^n$. 
Only then do we infer its domain, that is, the state preparation map.
This yields robustness to \emph{state preparation or measurement} (SPOM) errors $S$ or $M$ in the data model  (cf.\ \cref{eq:spam time-evolution data} of the main text). Both ways of denoising the signal from coherent and incoherent noise are crucial for our reconstruction method to be practically applicable. 

In the following, we detail algorithmic strategies for the different steps of our identification algorithm, relying on advanced state-of-the-art signal processing techniques that are capable of exploiting the entire structure of the problem, as well as developing new tools tailored to this specific structure.
For the sake of clarity we first consider the ideal data model \eqref{eq:data model}, and then discuss the effect and removal of SPAM errors in the algorithm. 

We begin in \cref{sec:frequency reconstruction} by introducing a novel super-resolving and denoising algorithm \emph{tensorESPRIT} that
is able to scalably resolve arbitrary degenerate frequency spectra of Hermitian matrices from samples of the corresponding one-parameter subgroup of the unitary group.
Given the eigenfrequencies, in \cref{sec:basis reconstruction} we discuss and compare different ways of reconstructing the eigenvectors of $h$ in the presence of constraints. 
In \cref{sec:spam} we then discuss the SPAM errors that affect the ideal data, and elaborate ways of partially removing those errors, giving rise to a robust recovery algorithm. 
In \cref{sec:full algorithm} we summarize the entire algorithm, before benchmarking its performance in \cref{sec:numerics}. 
Finally, in \cref{sec:error estimation} we explain how we obtain the systematic and statistical error bars for the plots in the main text. 

\section{Frequency extraction via rotational invariance}
\label{sec:frequency reconstruction}

The data of the form as elaborated upon in \eqref{eq:diagonalized data} consists of $N^2$ time series, each described by a linear combination of $N$ sinusoids oscillating at the eigenfrequencies of the Hamiltonian, which we wish to recover.
A simple approach to recover these frequencies would be to use Fourier analysis. This approach is limited by different imperfections since some of the Fourier coefficients can become small.
Furthermore, it is far from obvious how to combine the spectra recovered from each time series, especially if the spectrum is degenerate or nearly degenerate.
Additionally, Fourier approaches are typically limited in precision by Shannon's sampling theorem \cite{Shannon:1949}. 
Since we need the spectrum very accurately in order for the eigenvector reconstruction to converge, we need a more sophisticated approach.

To arrive at such a method, we make two structural observations. 
Firstly (\emph{sparsity}), we notice that the spectrum is sparse in Fourier space; there are exactly $N$ frequencies to be recovered.
This is a significantly more constraining structure than Shannon's band limitation alone.
Secondly (\emph{rotational-invariance}), the matrix time series $y_{m,n}[l]$ are noisy samples from the one-parameter subgroup of the unitary group $\mathrm{U}(N)$ generated by $h$ and parameterised by time. 

It will be useful to first think about the scalar ($N=1$) case where powerful algorithms exist, before extending those approaches to the matrix case with $N>1$.
The problem of extracting sparse frequencies (or also decay parameters, i.e. complex poles) from discrete scalar time series has been well studied for centuries \cite{Prony:1795}.
Modern stable algorithms for finding solutions to this problem have been devised in the context of direction-of-arrival estimation in array signaling \cite{schmidt_multiple_1986, royESPRITestimationSignalParameters1989}.
Here the goal is to find directions of arrival of a sparse set of electromagnetic signals from noisy snapshots detected by an array of antennas in the far field. 
Exploiting \emph{sparsity} and \emph{rotational $\mathrm{U}(1)$ invariance}, these methods utilize signal space estimation, denoising and low-rank Hankel structure to achieve precision beyond Shannon's theorem, which is a phenomenon called \emph{super-resolution}.
Theoretical understanding of super-resolution has only been developed recently \cite{CandesFernandez-Granda:2013, candes_towards_2014}.

\subsection{ESPRIT} \label{ssec:scalar esprit}

A particularly elegant example, relying solely on linear algebra routines, is the ESPRIT algorithm proposed in Ref. \cite{RoyPaulrajKailath:1986} and analysed in Refs. \cite{Fannjiang:2016:ESPRIT, LiLiaoFannjiang:2019:Super-resolution}. 
Let $t_l = l\Delta t$ be equally spaced with $l \in \{0,\dots,L\}$.
ESPRIT then assumes time series data of the form
\begin{equation}
\label{eq:scalar timeseries esprit}
    x[l] = \sum_{k = 1}^N x_k \e^{-i\lambda_k l \Delta t} = \sum_{k=1}^N x_k z_k^l,
\end{equation}
where $z_k \coloneqq \e^{-i\lambda_k\Delta t}$. 
The first step of the ESPRIT algorithm is to form the \emph{Hankel matrix} 
\begin{equation}
    \Hankel_K(x) = \begin{pmatrix}
      x[0] & x[1] & \dots & x[L-K] \\
      x[1] & x[2] & \dots\\
      \vdots &&\ddots\\
      x[K] & x[K+1] &\dots & x[L]
    \end{pmatrix},
\end{equation}
of the scalar time series \eqref{eq:scalar timeseries esprit},  defined for an integer $0 < K < L$.
The Hankel matrix admits Vandermonde decomposition
\begin{equation}\label{eq:vandermonde_config}
    \Hankel_K(x) = \Psi^K(z) \diag(x) \left(\Psi^{L-K}(z)\right)^T,
\end{equation}
where the Vandermonde matrix $\Psi^J(.)$ for an integer $J$ is given by
\begin{equation}
    \Psi^J(z) \coloneqq \begin{pmatrix}
      1&1&\dots&1\\
      z_1&z_2&\dots&z_N\\
      z_1^2&z_2^2&\dots&z_N^2\\
      \vdots&\vdots&\ddots&\vdots\\
      z_1^J & z_2^J & \dots & z_N^J
    \end{pmatrix} \in \mathbb{C}^{(J+1)\times N}.
\end{equation}
This decomposition makes the rotational invariance of the data apparent.
To see this, notice that inserting $\diag(z)$ between the first and second matrix in \eqref{eq:vandermonde_config} of the decomposition gives rise to a Hankel matrix of shifted data $x[l]$ for $l=1,\dots,L+1$.

Denote by $\Hankel_K^{\downarrow(\uparrow)}(x) \in \mathbb{C}^{K\times (L-K+1)}$ the matrices obtained from $\Hankel(x)_K \in \mathbb{C}^{(K+1)\times(L-K+1)}$ by deleting the last (first) row. 
We can write
\begin{equation}\label{eq:esprit_decomp}
\begin{split}
    \Hankel_K^\downarrow(x) &= \Psi^{L-1}(z)\diag(x)\left(\Psi^{L-K}(z)\right)^T\,,\\
    \Hankel_K^\uparrow(x) &= \Psi^{K-1}(z)\diag(z)\diag(x)\left(\Psi^{L-K}(z)\right)\,1.
\end{split}
\end{equation}
Note that $\diag(z),\diag(x) \in \mathbb{C}^{N\times N}$. 
It can be shown that the Vandermonde matrices have full row rank if the frequencies are non-degenerate and $N+1\le K \le L-N$.
Hence, in this case, we have 
\begin{align}\rank\left(\Hankel_K^{\downarrow(\uparrow)}(x)\right) = \rank\left(\Hankel_K(x)\right) = N\,. 
\end{align}
We can, thus, denoise the data by using the rank-$N$ approximation of $\Hankel_K(x)$, 
\begin{align}\label{eq:rankNapproximation}
\widehat{\Hankel}_K(x) \coloneqq U_{[N]}\Sigma_{[N]}V^\dag_{[N]}, 
\end{align}
where $U\Sigma V^\dag$ is the \emph{singular value decomposition} (SVD) of $\Hankel_K(x)$ and the subscript $[N]$ denotes restriction to the subspace of the $N$ dominant singular values. 
From the denoised Hankel matrix we can obtain the denoised shifted matrices $\widehat{\Hankel}^{\downarrow(\uparrow)}_K(x)$.

The decompositions \eqref{eq:esprit_decomp} together with the rank considerations imply that
\begin{equation}
    \Psi \coloneqq \left(\widehat{\Hankel}_K^\downarrow(x)\right)^+ \widehat{\Hankel}_K^\uparrow(x),
\end{equation} 
where $(\cdot)^+$ denotes the pseudoinverse, has in absence of noise non-vanishing eigenvalues $\{z_1,\dots,z_N\}$.
The eigenvalues of $\Psi$ therefore provide an estimate of the eigenfrequencies $\{\lambda_k\}_{k=1}^N$, if the time step $\Delta t$ is chosen such that there exists a branch cut of the complex logarithm that uniquely identifies $\lambda_k$ from $z_k$ for each $k$.

In order to apply the ESPRIT algorithm to the data of the Hamiltonian identification problem, we compute a suitable scalar time series from the matrix-valued time series \eqref{eq:data model}. 
Optimal performance of ESPRIT in resolving the spectrum is expected when the Fourier coefficients of the time-series are as large as possible and of the same order of magnitude \cite{LiLiaoFannjiang:2019:Super-resolution}. 
This is particularly important for large $N$ with unavoidably many adjacent frequencies.
In the data model \eqref{eq:data model}, we can achieve this by taking the matrix trace at each time
\begin{equation}\label{eq:ideal_ESPRIT_input}
    F[l] = \Tr \left[ y[l]\right] = \frac{1}{2} \sum_{k=1}^N \e^{-i \lambda_k l \Delta t}\, .
\end{equation}

In practice, imperfections in the state-preparation and measurements (due to finite speed of control) will alter the data model \eqref{eq:data model}. 
We describe a pre-processing procedure to still get a signal of approximately the form~\eqref{eq:ideal_ESPRIT_input} in the presence of SPOM in \cref{ssec:sperror}. 

The time complexity of ESPRIT is dominated by calculating the SVD which for an $m\times n$ matrix requires $\mc O(\min\{m^2n, mn^2\})$ flops, e.g., using Householder reflections \cite{GolubLoan}.
For the Hankel dimension $K \in \mc O(N)$ ESPRIT, thus, takes time $\mc O(N^2 L)$, with $L$ the length of the time series and $N$ the number of sinusoids.
Memory cost stays of the order of the size of the input $\mc O(N^2 L)$.
  
\begin{figure}[tb]
 \begin{algorithm}[H]
   \caption{$\operatorname{ESPRIT}(F, K)$ (frequency extraction)}\label{alg:ESPRIT}
   \begin{algorithmic}[1]
     \Require $F \in \CC^{(L+1)}$, $K \leq L$.
     \State Set $H = \Hankel_K(F)$.
     \State Calculate the SVD of $H = U \Sigma V^\dagger$.
     \State Set $P_\text{signal} = U_{[N]} (U_{[N]})^\dagger$. 
     \State Calculate $\Psi = (P_\text{signal} H^{\uparrow})^+ H^{\downarrow}$.
     \State Calculate $z = \operatorname{eigenvalues}(\Psi)$. 
     \Ensure $z$.
   \end{algorithmic}
 \end{algorithm}
\end{figure}

\subsection{Tensor ESPRIT}
\label{ssec:tensor esprit}

ESPRIT can, in principle, recover frequencies that are considerably closer together than the Shannon-Nyquist rate. 
For increasing system sizes $N$ the spectrum becomes unavoidably closer and closer to being degenerate (if one does not increase the energy scale extensively with the system size).
For this reason, for sufficiently large $N$ the approach of reducing the data to a single scalar time-series and running ESPRIT will inevitably fail.
We here develop a novel algorithm that directly works on the matrix time series and makes use of the entire $\mathrm U(N)$-rotational invariance of the data. 
To this end, we utilize tensor-network methods, carefully exploiting the properties of the data series over different contractions in the tensor network. 
The new \emph{tensorESPRIT} algorithm is capable of resolving even degenerate Hamiltonian spectra and, thus, enables finding the complete eigenfrequency spectrum even of large-size instances.

With $t_l = l\Delta t$ equally spaced with $l\in\{0,\dots,L\}$, the time series \eqref{eq:diagonalized data} takes the form
\begin{equation}
    y[l] = \frac{1}{2}\sum_{p=1}^N z_p^l \proj{v_p},
\end{equation}
where again $z_p \coloneqq e^{-i\Delta t\lambda_p}$.
Let $Q = \sum_p \ket {v_p} \bra p$ be an orthogonal matrix with columns $\ket{v_p}$.
We form the Hankel tensor, which we define to be
\begin{equation}
    \Hankel_K(y)_{m,k,l,n} \coloneqq 2 y_{m,n}[k+l] = \sum_{p=1}^N z_p^{k+l}Q_{m,p} Q_{n,p}
\end{equation}
with $k\in\{0,\dots,K\}$ and $l\in\{0,\dots,L-K\}$.
Due to rotational invariance of the data, $\Hankel_K(y)$ admits a tensor version of the Vandermonde decomposition.
In tensor network notation\footnote{In this notation, a degree $k$ tensor is represented by a box with $k$ legs attached to it, representing the indices. If two boxes are connected by a leg, the corresponding index is contracted over. See Ref.~\cite{bridgemanHandwavingInterpretiveDance2017} for a thorough introduction to this notation.} we find
\begin{equation}\label{eq:Hankel_tensor_decomposition}
    \begin{tikzpicture}[
      squarednode/.style={rectangle, draw=blue!70, fill=blue!5, thick, minimum size = 7mm},
        ]   

      \node[squarednode] (alpha) {$Q$};
      \node[squarednode] (phi_1) [right=0.4cm of alpha] {$\Phi^{K}$};
      \node[squarednode] (phi_2) [right=0.4cm of phi_1] {$\Phi^{L-K}$};
      \node[squarednode] (beta) [right=0.4cm of phi_2] {$Q^T$};
        \node (i) [below=0.5cm of alpha] {$m$};
        \node (k) [below=0.5cm of phi_1] {$k$};
        \node (l) [below=0.5cm of phi_2] {$l$};
        \node (j) [below=0.5cm of beta] {$n$};
        \node (eq) [left=2mm of alpha] {$=$};
        \node[squarednode] (lhs) [left=2mm of eq] {\mbox{\quad$\Hankel_K(y)$\quad}};
        \node (ref) [below=0.5 of lhs] {};
        \node (i2) [left=0.7 of ref.base] {$m$};
        \node (j2) [right=0.7 of ref.base] {$n$};
        \node (k2) [left=0.2 of ref.base] {$k$};
        \node (l2) [right=0.2 of ref.base] {$l$};
        \node (i3) [above=0.75 of i2.base] {};
        \node (j3) [above=0.7 of j2.base] {};
        \node (k3) [above=0.75 of k2.base] {};
        \node (l3) [above=0.75 of l2.base] {};
        \node (comma) [right=0mm of beta] {,};
      
        \draw (i.north) -- (alpha.south);
        \draw (k.north) -- (phi_1.south);
        \draw (l.north) -- (phi_2.south);
        \draw (j.north) -- (beta.south);
        \draw (alpha.east) -- (phi_1.west);
        \draw (phi_1.east) -- (phi_2.west);
        \draw (phi_2.east) -- (beta.west);
        \draw [shorten >=0.5mm] (i2.north) -- (i3.south);
        \draw (j2.north) -- (j3.south);
        \draw (k2.north) -- (k3.south);
        \draw (l2.north) -- (l3.south);
    \end{tikzpicture}
\end{equation}
where
$\Phi^J$ for an integer $J$ is the Vandermonde tensor with components
\begin{equation}
    \begin{tikzpicture}[
          squarednode/.style={rectangle, draw=blue!70, fill=blue!5, very thick, minimum size = 7mm},
          ]

        \node[squarednode] (phi) {$\Phi^J$};
        \node (i) [left=0.3cm of phi] {$i$};
        \node (j) [right=0.3cm of phi] {$j$};
        \node (k) [below=0.3cm of phi] {$k$};
        \node (rhs) [right=1mm of j] {$\coloneqq \delta_{i,j} z_i^k, \quad k \in \{0,\dots,J\}.$};

        \draw (k.north) -- (phi.south);
        \draw (phi.east) -- (j.west);
        \draw (i.east) -- (phi.west);
    \end{tikzpicture}
\end{equation}
We will obtain an unfolding $\mathcal{H}_K(y) \in \mathbb{C}^{(K+1)N \times (L-K+1)N}$ of $\Hankel_K(y)$ by grouping together the indices $m,k$ and $l,n$, respectively, such that the $N\times N$ block structure of $\mathcal{H}_K(y)$ is
\begin{equation}\label{eq:HankelMatrication}
    \mathcal{H}_K(y) = 
    \begin{pmatrix}
      y[0] & y[1] & \dots & y[L-K] \\
      y[1] & y[2] & \dots\\
      \vdots &&\ddots\\
      y[K] & y[K+1] &\dots & y[L]
    \end{pmatrix}.
\end{equation}
Decomposition \eqref{eq:Hankel_tensor_decomposition} implies that $\mathcal{H}_K(y)$ can be written as the product of an $NK \times N$ matrix, the correspondingly unfolded $Q^T\Phi^K$, and an $N\times N(M-L)$ matrix, the unfolded $\Phi^{L-K}Q$.
Hence the rank of $\mathcal{H}_K(y)$ is upper bounded by $N$, which allows us to denoise the data by applying a rank-$N$ approximation to $\mathcal{H}_K(y)$, defined in \eqref{eq:rankNapproximation}. 

Similarly to ESPRIT, we define shifted tensors
\begin{equation}
\begin{split}
\Hankel_K^\downarrow(y)_{m,k,l,n} &\coloneqq \Hankel_K(y)_{m,k,l,n},\\
\Hankel_K^\uparrow(y)_{m,k,l,n} &\coloneqq \Hankel_K(y)_{m,k+1,l,n} 
\end{split}
\end{equation}
with the restriction $k\in{0,\dots,K-1}$ in both cases.
These tensors admit Vandermonde decompositions
\begin{equation}
    \begin{tikzpicture}[
          squarednode/.style={rectangle, draw=blue!70, fill=blue!5, very thick, minimum size = 7mm},
          redsquarednode/.style={rectangle, draw=red!70, fill=red!5, very thick, minimum size = 7mm},
          ] 

        \node[squarednode] (alpha) {$Q$};
      \node[squarednode] (phi_1) [right=0.3cm of alpha] {$\Phi^{K-1}$};
      \node[squarednode] (phi_2) [right=0.3cm of phi_1] {$\Phi^{L-K}$};
      \node[squarednode] (beta) [right=0.3cm of phi_2] {$Q^T$};
        \node (i) [below=0.5cm of alpha] {};
        \node (k) [below=0.5cm of phi_1] {};
        \node (l) [below=0.5cm of phi_2] {};
        \node (j) [below=0.5cm of beta] {};
        \node (eq) [left=2mm of alpha] {$=$};
        \node[squarednode] (lhs) [left=2mm of eq] {\mbox{\quad$\Hankel_K^\downarrow(y)$\quad}};
        \node (ref) [below=0.5 of lhs] {};
        \node (i2) [left=0.7 of ref.base] {};
        \node (j2) [right=0.7 of ref.base] {};
        \node (k2) [left=0.2 of ref.base] {};
        \node (l2) [right=0.2 of ref.base] {};
        \node (i3) [above=0.5 of i2] {};
        \node (j3) [above=0.5 of j2] {};
        \node (k3) [above=0.5 of k2] {};
        \node (l3) [above=0.5 of l2] {};
      
        \draw (i.north) -- (alpha.south);
        \draw (k.north) -- (phi_1.south);
        \draw (l.north) -- (phi_2.south);
        \draw (j.north) -- (beta.south);
        \draw (alpha.east) -- (phi_1.west);
        \draw (phi_1.east) -- (phi_2.west);
        \draw (phi_2.east) -- (beta.west);
        \draw (i2.north) -- (i3.south);
        \draw (j2.north) -- (j3.south);
        \draw (k2.north) -- (k3.south);
        \draw (l2.north) -- (l3.south);
      \end{tikzpicture}
\end{equation}
\begin{equation}
    \begin{tikzpicture}[
          squarednode/.style={rectangle, draw=blue!70, fill=blue!5, very thick, minimum size = 7mm},
          redsquarednode/.style={rectangle, draw=red!70, fill=red!5, very thick, minimum size = 7mm},
          ] 

      \node[squarednode] (alpha) {$Q$};
      \node[redsquarednode] (lambda) [right=0.3cm of alpha] {$\Lambda$};
      \node[squarednode] (phi_1) [right=0.3cm of lambda] {$\Phi^{K-1}$};
      \node[squarednode] (phi_2) [right=0.3cm of phi_1] {$\Phi^{L-K}$};
      \node[squarednode] (beta) [right=0.3cm of phi_2] {$Q^T$};
        \node (i) [below=0.5cm of alpha] {};
        \node (k) [below=0.5cm of phi_1] {};
        \node (l) [below=0.5cm of phi_2] {};
        \node (j) [below=0.5cm of beta] {};
        \node (eq) [left=2mm of alpha] {$=$};
        \node[squarednode] (lhs) [left=2mm of eq] {\mbox{\quad$\Hankel_K^\uparrow(y)$\quad}};
        \node (ref) [below=0.5 of lhs] {};
        \node (i2) [left=0.7 of ref.base] {};
        \node (j2) [right=0.7 of ref.base] {};
        \node (k2) [left=0.2 of ref.base] {};
        \node (l2) [right=0.2 of ref.base] {};
        \node (i3) [above=0.5 of i2] {};
        \node (j3) [above=0.5 of j2] {};
        \node (k3) [above=0.5 of k2] {};
        \node (l3) [above=0.5 of l2] {};
        \node (comma) [right=0mm of beta] {,};
      
        \draw (i.north) -- (alpha.south);
        \draw (k.north) -- (phi_1.south);
        \draw (l.north) -- (phi_2.south);
        \draw (j.north) -- (beta.south);
        \draw (alpha.east) -- (lambda.west);
        \draw (lambda.east) -- (phi_1.west);
        \draw (phi_1.east) -- (phi_2.west);
        \draw (phi_2.east) -- (beta.west);
        \draw (i2.north) -- (i3.south);
        \draw (j2.north) -- (j3.south);
        \draw (k2.north) -- (k3.south);
        \draw (l2.north) -- (l3.south);
      \end{tikzpicture}
\end{equation}
where $\Lambda = \diag(z)$.
Now we make the crucial observation that in the noiseless case for any fixed $k,l$ the matrix
\begin{equation}\label{eq:tesprit_crucial_observation}
    \begin{tikzpicture}[
          squarednode/.style={rectangle, draw=blue!70, fill=blue!5, very thick, minimum size = 7mm},
          ]
        \node[squarednode] (U) at (0,0) {$U^{(k,l)}$};
        \node (iU) [left=0.3 of U] {};
        \node (jU) [right=0.3 of U] {};
        \node (def) [right=0.1 of jU] {$\coloneqq$};
        \node (i) [right=0. of def] {};
        \node[squarednode] (phi1) [right=0.3 of i] {$\Phi^{K-1}$};
        \node[squarednode] (phi2) [right=0.3 of phi1] {$\Phi^{L-K}$};
        \node[squarednode] (beta) [right=0.3 of phi2] {$Q^T$};
        \node (k) [below=0.3 of phi1] {$k$};
        \node (l) [below=0.3 of phi2] {$l$};
        \node (j) [right=0.3 of beta] {};
        
        \draw (iU.east) -- (U.west);
        \draw (U.east) -- (jU.west);
        \draw (i.east) -- (phi1.west);
        \draw (phi1.south) -- (k.north);
        \draw (phi1.east) -- (phi2.west);
        \draw (phi2.south) -- (l.north);
        \draw (phi2.east) -- (beta.west);
        \draw (beta.east) -- (j.west);
    \end{tikzpicture}
\end{equation}
is unitary. 
Furthermore $U^{(k,l)} = U^{(k',l')}$ whenever $k+l=k'+l'$.
Choosing $k,l,k',l'$ such that $k+l=k'+l'$, we compute the matrix
\begin{equation}\label{eq:tensorESPRIT_robust_contraction}
    \begin{tikzpicture}[
          squarednode/.style={rectangle, draw=blue!70, fill=blue!5, very thick, minimum size = 7mm},
        redsquarednode/.style={rectangle, draw=red!70, fill=red!5, very thick, minimum size = 7mm},
        dashednode/.style={rectangle,draw=black,dashed,minimum width = 2.3cm,minimum height = 2cm},
          ]

        \node (i) at (0,0) {};
        \node[squarednode] (H2) [right=4mm of i] {$\Hankel_K^\uparrow(y)$};
        \node[squarednode] (H1) [right=5mm of H2] {$\Hankel_K^\downarrow(y)$};
        \node[dashednode] (box) [above=-1.2cm of H1.base] {};
        \node (pseudoinverse) [above right=-5mm of box] {+};
        \node (j) [right=4mm of H1] {};
        \node (ref1) [below=0.5cm of H1] {};
        \node (ref2) [below=0.5cm of H2] {};
        \node (k) [left=0.3cm of ref2] {$k$};
        \node (l) [right=0.3cm of ref2] {$l$};
        \node (k') [left=0.3cm of ref1] {$k'$};
        \node (l') [right=0.3cm of ref1] {$l'$};
        \node (ktop) [above=0.4cm of k] {};
        \node (ltop) [above=0.4cm of l] {};
        \node (k'top) [above=0.4cm of k'] {};
        \node (l'top) [above=0.4cm of l'] {};
        \node (comma) [right=-0.3 of j] {,};

        \draw (i.east) -- (H2.west);
        \draw (k.north) -- (ktop.south);
        \draw (l.north) -- (ltop.south);
        \draw (k'.north) -- (k'top.south);
        \draw (l'.north) -- (l'top.south);
        \draw (H2.east) -- (H1.west);
        \draw (H1.east) -- (j.west);

        \node (llhs) [left=-0.2 of i] {$\coloneqq$};
        \node[squarednode] (A) [left=0.3 of llhs] {$A^{(k,l,k',l')}$};
        \node (iA) [left=0.3 of A] {};
        \node (jA) [right=0.3 of A] {};
        
        \draw (iA.east) -- (A.west);
        \draw (jA.west) -- (A.east);
    \end{tikzpicture}
\end{equation}
where the box with the plus denotes that we fix all the indices inside the box and then take the psuedoinverse of the matrix indexed by legs going through the box.
From the observation \eqref{eq:tesprit_crucial_observation} we conclude that (in the absence of noise) $A^{(k,l,k',l')} = Q^T \Lambda Q$ has spectrum $\{z_i\}_{i=1}^n$. 

It may seem that we could have used the much cheaper Hermitian conjugate instead of the pseudoinverse in \eqref{eq:tensorESPRIT_robust_contraction}. But in the presence of noise, $U^{(k,l)}$ can deviate from being unitary, and hence using the pseudoinverse leads to a more accurate recovery.
In fact, tensorESPRIT with the pseudoinverse can be applied more generally even if $Q$ and $Q^T$ are replaced by two (potentially different) invertible matrices, such as in the case of our SPAM model \eqref{eq:spam data model}.

Combining the spectra of matrices $A^{(k,l,k',l')}$ for different choices of $k,k',l,l'$ improves the noise robustness of the estimate.
To this end, we define a set $\mathcal{S}$ of tuples $(k,l,k',l')$ with $k+l=k'+l'$, and compute
\begin{equation}
    \hat A = \frac{1}{|\mathcal{S}|}\sum_{(k,l,k',l') \in \mathcal{S}} A^{(k,l,k',l')}\,.
\end{equation}
The spectrum of $\hat A$ is the arithmetic mean of the spectra of $A^{(k,l,k',l')}$ for $(k,l,k',l') \in \mathcal{S}$ and yields a robust estimate of $\{z_i\}_{i=1}^n$.
We find empirically that 
\begin{equation}
\begin{split}\label{eq:tesprit:stdS}
    \mathcal{S} = \{(k,l,k',l'):\ 0 \leq k = k' < K,\  0 \leq l = l' \leq L-K\}
\end{split}
\end{equation}
gives good performance  and we use this definition throughout this manuscript.

As noted above, tensorESPRIT is robust against non-singular $S$ and $M$ in the SPAM data model \eqref{eq:spam data model}. Poor conditioning of these matrices however still reduces its accuracy.
Hence, we can apply tensorESPRIT to the data $y[l]$ without the pre-processing (\cref{ssec:sperror}) required for ESPRIT.
The tensorESPRIT algorithm is summarized as \cref{alg:tensorESPRIT}.

The time complexity of tensorESPRIT is $\mathcal{O}((LN^2 + | \mathcal{S} |) \times N^3)$ flops, while memory complexity is $\mathcal{O}(LN^2)$---of the order of the size of the input, using implicit definition of the Hankel tensor.
The first term in the time complexity comes from the SVD of the unfolded Hankel tensor, which with the choice $K \in \mc O(N)$ is an $\mc O(N^2) \times \mc O(LN)$ matrix, where we expect $L\gg N$. 
Using randomized truncated SVD methods \cite{halkoFindingStructureRandomness2010}, this term can be further improved to $\mc O(LN^3 \log(N)$.
The second term comes from the computation of the matrix $\hat A$ from $\widehat{\mathcal{H}}$. 
The maximal size of $\mathcal{S}$ is $\mathcal{O}(L^3)$, but the set $\mathcal{S}$ used in this work has size in $\mathcal{O}(L^2)$.

\begin{figure}[tb]
 \begin{algorithm}[H]
   \caption{$\operatorname{tensorESPRIT}(y, K, \mathcal{S})$ (frequency extraction)}\label{alg:tensorESPRIT}
   \begin{algorithmic}[1]
     \Require $y \in \CC^{N \times N \times (L+1)}$,
     $K \leq L$,
     $\mathcal{S}$.
     \State Set $\mathcal{H}_K$ to be the unfolding \eqref{eq:HankelMatrication} of $\Hankel_K(y)$.
     \State Set $\widehat {\mathcal{H}} =  U_{[N]}\Sigma_{[N]}V^\dag_{[N]} $ as the rank-$N$ approximation of $\mathcal{H}_K$.
     \State From $\widehat{\mathcal{H}}$ as Hankel tensor, calculate $A^{(k,l,k',l')}$ of \eqref{eq:tensorESPRIT_robust_contraction}  for all $(k,l,k',l') \in \mathcal{S}$.
     \State Set $\hat A = |\mathcal{S}|^{-1} \sum_\mathcal{S} A^{(k,l,k',l')}$.
     \State Calculate $z = \operatorname{eigenvalues}(\hat A)$. 
     \Ensure $z$.
   \end{algorithmic}
 \end{algorithm}
\end{figure}


\section{Reconstruction of eigenvectors}
\label{sec:basis reconstruction}

After extracting the eigenfrequencies $\{\lambda_k\}$, in the second step of the Hamiltonian identification, we infer the eigenvectors of $h$.  
Let $\Omega$ be the support set of the matrix $h$ of the Hamiltonian model as  determined by the interaction graph of $h$. 
This set comprises of index pairs $(i,j)$ corresponding to a hopping term between sites $i$ and $j$.
For example, for the non-interacting Bose-Hubbard model (\cref{eq:bose-hubbard H} of the main text) with nearest neighbour interaction, the support is given by the diagonal and first-order off-diagonal terms. For a support set $\Omega$, $\bar\Omega$ denotes the complement of its support.
For a matrix $X$, $X_\Omega$ denotes the sub-matrix restricted to the entries in $\Omega$ and $\|X\|_F = \sum_{i,j} |X_{i,j}|^2$ the Frobenius norm. 

The eigenvector recovery task can be formulated as the least-squares optimization problem
\begin{equation}\label{eq:generalEigReconstruction}
\begin{split}
  \operatorname*{minimise}_{\{\ket{v_i}\}}&\quad \sum_l \left\| y[l] - \sum_k \e^{-\i \lambda_k t_l} \ketbra{v_k}{v_k}\right\|^2_F , \\
  \operatorname{subject\ to}&\quad \braket{v_i}{v_j}= \delta_{i,j}, \quad \left(\sum_k \lambda_k \ketbra{v_k}{v_k}\right)_{\bar\Omega} = 0. 
\end{split}\end{equation}
The objective function is a quartic polynomial. 
In addition, we encounter a non-convex constraint enforcing the orthogonality and quadratic support enforcing the sparsity. 

To simplify the expressions it is helpful to introduce the following notation. 
Let $\operatorname{vec}: \CC^{d\times d'}  \to \CC^{d\cdot d'}$ denote row-wise vectorization, which acts as  $\ketbra i j \mapsto \ket i \ket j$ on the orthogonal basis.  
Using row-wise vectorization, we rewrite the data $y[l] \in \CC^{N \times N}$ with $l \in [L]$ as a single $(L+1) \times N^2$ matrix 
\begin{equation}
  Y = \begin{pmatrix}
      \operatorname{vec}(y[0])^T \\ \operatorname{vec}(y[1])^T \\ \vdots \\ \operatorname{vec}(y[L])^T
    \end{pmatrix}\, .
\end{equation}

Let $Q = \sum_k \ket {v_k} \bra k$ be an orthogonal matrix with columns $\ket{v_k}$. 
We define the map
$\Pi: \O(N) \to \RR^{N\times N^2}$ on the orthogonal group $\O(N)$ as 
\begin{equation}
  \Pi(Q) \coloneqq \sum_{k} \ket{k} \bra{v_k}\bra{v_k}\, . 
\end{equation}
This definition is equivalent with
\begin{equation}\label{eq:PiCoordinateExpression}
      \bra k\Pi(Q)\ket l\ket m = \Pi(Q)_{k,(l,m)} = Q_{l,k}Q_{m,k}\, .
\end{equation} 
Furthermore, let $A^\lambda$ be the matrix with the time series $l \mapsto \e^{-\i \lambda_k t_l}$ as its $k$-th column.  
For equidistant times, $A^\lambda$ is the previously encountered Vandermonde matrix.
Lastly, we define $H^\lambda: \RR^{N\times N^2} \to \RR^{N\times N}$ that given $\Pi(Q)$ returns the Hamiltonian matrix associated to $Q$ with eigenvalues $\lambda$.  
Explicitly, $H^\lambda(\Pi) = \operatorname{vec}^{-1} ( \sum_i \bra i \diag(\lambda) \Pi)$. 

In this notation the optimization problem \eqref{eq:generalEigReconstruction} can be recast more compactly as 
\begin{equation}\label{eq:generalEigReconstruction2}
\begin{split}
  \operatorname*{minimise}_{Q \in \O(N)}&\quad \left\| Y - A^\lambda \Pi(Q)\right\|^2_F , \\
  \operatorname{subject\ to}&\quad  (H^\lambda(\Pi(Q)))_{\bar\Omega} = 0\, . 
\end{split}\end{equation}

In the following, we detail two distinct strategies for solving the optimization problem \eqref{eq:generalEigReconstruction2}.
First, we solve the problem for $\Pi(Q)$ using linear inversion in \cref{ssec:linear inv postprojection}. We can further enforce the projector structure of the rows of $\Pi(Q)$ by post-projection, but not their orthogonality.
Second, we make use of gradient-descent methods constrained to the non-convex manifold of orthogonal matrices in \cref{ssec:nonconvex}. 
We approximately incorporate the locality constraint by regularization.

\subsection{Linear inversion with post-projection}
\label{ssec:linear inv postprojection}

Of course, knowing the eigenspace projectors $\Pi(Q)$ and $\lambda$ is already sufficient to calculate the 
corresponding Hamiltonian matrix.  
The simplest approach to the optimization problem is thereby to neglect the exact dependency of $\Pi$ on $Q$, the 
 non-convex constraint and the support constraint.
This yields the significantly simpler optimization problem 
\begin{equation}\label{eq:linearEigReconstruction2}
\begin{split}
  \operatorname*{minimise}_{\Pi \in \CC^{N\times N^2}}&\quad \left\| Y - A^\lambda \Pi\right\|^2_F 
\end{split}\end{equation}
where we slightly overloaded the symbol $\Pi$.  
The linear inverse problem 
\eqref{eq:linearEigReconstruction2}
can be solved in closed form as 
\begin{align}
\label{eq:pseudo inverse}
  \Pi &= (A^\lambda)^+ Y\,. 
\end{align}
Note that in principle it is also straight-forward to solve for the linear support constraint when setting-up the optimization problem.

This matrix will in general not retain the projector structure of the signal, however. Indeed, recall that $\Pi = \Pi(Q)$ with $Q = \sum_l \ket l \bra {v_l} \in \O(N)$, so the $l$-th row of $\Pi$ is the vectorization of the projector onto the eigenspaces spanned by $\ket {v_l}$.  
In order to enforce this structure, we make use of a post-projection step:
Given a matrix $P$, we can project it onto the manifold of real unit-rank projectors in order to enforce this constraint. 
To achieve this, we project them to Hermitian matrices $P' = (P + P^\dagger)/2$, perform an eigenvalue decomposition $P' = U\Lambda U^\dagger$, select the eigenvector $U_1$ of the absolutely largest eigenvalue and calculate $P'' = \Re\{U_1 U_1^\dagger\}$.  
We summarize the corresponding algorithm in \cref{alg:linInvPP}.
Notice that the resulting projectors will in general not be mutually orthogonal. 

The run-time of the algorithm scales as $\mc O(LN^2 + N^3)$, where the first term comes from the 
pseudo-inversion and the second term from performing $N$ unit-rank projections of $N\times N$ matrices. 
Both steps can be implemented, e.g., using an SVD.

\begin{figure}[tb]
\begin{algorithm}[H]
  \caption{$\operatorname{linInvPP}(y, A^\lambda, \lambda)$}\label{alg:linInvPP}
  \begin{algorithmic}[1]
    \Require Data $y$, map $A^\lambda$.
    \State Calculate $\Pi = (A^\lambda)^+ Y$.
    \For{$k \in [N]$}
      \State Set $P_k = \Pi[k, :]$ and reshape to $N \times N$ matrix. 
      \State Project $P_k \leftarrow (P_k + P_k^\dagger)/2$
      \State Calculate $u_k$ the eigenvector to the largest eigenvalue of $P_k$.
      \State Set $P_k = U_kU_k^\dagger$.
      \State Vectorize $P_k$ and set $\Pi[k, :] = P_k$. 
    \EndFor
    \Ensure Projector matrix $\Pi$. 
  \end{algorithmic}
\end{algorithm}

\end{figure}

\subsection{Non-convex manifold optimization}
\label{ssec:nonconvex}

Taking the structure of the reconstruction problem for the eigenvectors \eqref{eq:generalEigReconstruction} seriously requires us to account for non-convex orthonormality constraints.  
In the following, we detail how the optimization problem with orthonormality constraints can be solved using geometrical optimization techniques over the manifold structure that the orthogonal group exhibits as a Lie group. 
See, e.g., Refs.~\cite{EdelmanAriasSmith:1998,absil_optimization_2009} for a general introduction and a further reference, and Refs.~\cite{luchnikov2020qgopt, luchnikov2020riemannian, HangleiterEtAl:2020, KrumnowEtAl:2016}  for manifold optimization in the context of quantum information and technologies. 

To this end, we first neglect the sparsity constraint in \eqref{eq:generalEigReconstruction2}, and consider the optimization problem
\begin{equation}\label{eq:nonConvexEigReconstruction}
\begin{split}
  \operatorname*{minimise}_{Q \in \O(N)}&\quad f_{y,A^\lambda}(Q) = \frac12 \left\| Y - A^\lambda \Pi(Q)\right\|^2_F\, . 
\end{split}\end{equation}
Many standard first-order and second-order optimization algorithm, such as gradient descent methods or Newton's method, readily generalize to matrix manifolds by using the differential structure and Riemannian geometry provided by a suitable embedding \cite{absil_optimization_2009}.  
In the following we regard $\O(N) \subset \RR^{N\times N}$ as a submanifold of the Euclidean space defined by its standard embedding as a matrix group.  
We now formulate a conjugate gradient algorithm for optimizing $f$ over $\O(N)$ proposed in Refs.~\cite{EdelmanAriasSmith:1998, abrudan_conjugate_2009}.  
The conjugate gradient algorithm iterates the following basic steps: (i) At a point $Q_k$ on the manifold, determine a search direction $V_k$ from the current gradient of the objective function $f$ and the conjugacy conditions to the previous search directions with respect to the Hessian of the objective function. (ii) Perform a line search to determine the next point $Q_{k+1}$ as the minimum of $f$ along a geodesic through $Q_k$ in direction $V_k$. 

Search direction and the gradient are elements of the tangent space of the manifold $\O(N)$. 
The tangent space of $\O(N)$ is given by $T_Q\O(N) = \{ V \in \RR^{N\times N} \mid V^T Q + Q^T V = 0 \}$ and can be equipped with the Riemannian metric 
\begin{equation}
\langle V, W \rangle_Q = \frac12 \Tr[VW^T], 
\end{equation}
$Q \in \O(N)$ and $V, W \in T_Q\O(N)$, which is induced by the Euclidean metric on the ambient space.  
The tangent space at the group identity $\Id \in \O(N)$ is given by skew-symmetric matrices and identified with the Lie algebra $o(N)$.  
For $V \in o(N) = T_{\Id} \O(N)$, we have that $\tilde V = V Q \in T_Q\O(N)$. 
The orthogonal projection onto $T_Q\O(N)$ with respect to the Euclidean metric is $\mc P_{T_Q\O(N)}: \RR^{n\times n} \to T_Q\O(N)$, 
\begin{equation}
Z \mapsto \frac12(Z - Q Z^T Q). 
\end{equation}

Given a search direction $V \in T_Q\O(N)$, a natural way to move forward on the manifold is along the geodesic of the Levi-Cevita connection defined by $\langle \cdot, \cdot \rangle_Q$.  
For $\O(N)$ a closed form of a geodesic $\gamma_{Q, \tilde V}$ in direction $\tilde V = V Q \in T_Q\O(N)$ through point $Q \in \O(N)$ is given by the matrix exponential, that here coincides with the exponential map from Lie theory, 
\begin{equation}\label{eq:geodesic}
  t \mapsto \gamma_{Q, \tilde V}(t) = \exp(V t) Q\, .
\end{equation}
Note that here $V \in o(N)$ instead of $\tilde V$ appears in the exponent.  
More generally, the notion of a retraction generalizes the idea of moving along the manifold in a search direction while still ensuring convergence of descent algorithms \cite{absil_optimization_2009}.  
Employing other retractions such as the Cayley transformation or projection using the QR-decomposition avoids the numerically costly matrix exponential and can reduce the computational complexity of the optimization algorithm.  
We will not pursue these alternatives here. 

To formulate the conjugacy condition between tangent vectors at different points of the manifold, we require the parallel transport of tangent vectors along geodesics.  
The vector $\tilde W = W Q \in T_Q\O(N)$ parallel transported along the geodesic $\gamma_{Q_, \tilde V}$, \eqref{eq:geodesic}, to $T_{\gamma_{Q, \tilde V}(t)}\O(N)$ is given by
\begin{equation}
   \Gamma^{t}_{\gamma_{Q,\tilde V}}(\tilde W) =  \e^{\frac12 V t }W\e^{\frac12 V t} Q\, . 
\end{equation}
For $\tilde W = \tilde V$ the direction of $\gamma_{Q, \tilde V}$ at $Q = \gamma_{\tilde V}(0)$ this expression reads
\begin{equation}
  \Gamma^{t}_{\gamma_{Q, \tilde V}}(\tilde V) = \tilde V Q^T \gamma_{Q, \tilde V}(t)\, .
\end{equation}

Ignoring the structure of the manifold and considering the standard embedding of $\O(N) \subset \RR^{n\times n}$,
 we can calculate the gradient with respect to the Euclidean metric of the ambient space. 
This Euclidean gradient can be subsequently projected onto the tangent space of the manifold $\O(N)$ to get a search direction of a gradient descent algorithm.  
The Euclidean gradient for our optimization problem is calculated as follows. We define 
\begin{equation}
g_{A^\lambda}(X) :=  \frac12 \left\|Y - A^\lambda X \right\|_F^2 
\end{equation}
and have $f_{y,A^\lambda} = g_{y,A^\lambda} \circ \Pi$. Then, by the chain rule, it holds that
\begin{equation}\label{eq:EuclideanGradient}
\begin{split}
  &(\nabla_E f_{y,A^\lambda}(Q))_{i,j} 
  = \left.\frac{\partial f_{y,A^\lambda}}{\partial Q_{i,j}}\right|_Q  \\
  &= \sum_{k,l,m} \left.\frac{\partial g_{A^\lambda}}{\partial X_{k,(l,m)}}\right|_{\Pi(Q)}\left.\frac{\partial \Pi_{k,(l,m)}}{\partial Q_{i,j}}\right|_Q.
\end{split}
\end{equation}
The outer derivative of the linear least-square problem is given by 
\begin{equation}\label{eq:EuclideanGradient:part1}
  \left.\frac{\partial g}{\partial X_{i,j}}\right|_{X} = 
  - \Re\{(A^{\lambda})^\dagger(Y - A^{\lambda}\cdot X)\}.
\end{equation}
The inner-derivative can be read-off from \eqref{eq:PiCoordinateExpression} to be
\begin{equation}\label{eq:EuclideanGradient:part2}
  \left.\frac{\partial \Pi_{k,(l,m)}}{\partial Q_{i,j}}\right|_Q = \delta_{i,l}\delta_{j,k}Q_{m,k} + \delta_{i,m}\delta_{j,k} Q_{l,k}\, .
\end{equation}
Note that carefully considering the order of the contractions of \eqref{eq:EuclideanGradient} and the sparsity pattern of the quantities, allows one to evaluate the gradient without performing routines in the full high-dimensional tensor spaces.  
At point $Q \in \O(N)$, we infer the Riemannian gradient via the tangent space projection as 
\begin{equation}
\begin{split}
\nabla f(Q) &= P_{T_{Q}\O(N)}[\nabla_E f(Q)] \\
&= \frac12[\nabla_E f(Q) - Q (\nabla_E f(Q))^T Q]\,.
\end{split}
\end{equation}
Given the previous search direction $H_{k-1}$ at point $Q_{k-1}$, the step size $t_{k-1}$, and the gradient $G_k$ at point $Q_k$, the new search direction is calculated as 
\begin{equation}
  H_k = - G_k + \gamma_k \hat H_{k-1}\, ,
\end{equation}
with $\gamma_k \in \RR$ and $\hat H_{k-1} = \Gamma^{t_k}_{\gamma_{Q_{k-1}, H_{k-1}}} H_{k-1}$, the previous search direction $H_{k-1}$ is parallel transported from $Q_{k-1}$ to $Q_k$.  
Exact conjugacy requires 
\begin{equation}
  \gamma_k = \frac{\operatorname{Hess} f_{y,A^{\lambda}} (G_k, \hat H_{k-1})}{\operatorname{Hess} f_{y,A^{\lambda}}(\hat H_{k-1}, \hat H_{k-1})}
\end{equation}
and can be approximated using the Polak-Ribière formula that arises from the finite difference approximation to the Hessian
\begin{equation}
  \gamma_k = \frac{
  \langle G_k - \hat G_{k-1}, G_k \rangle_{Q_k}
  }{
  \langle G_{k-1}, G_{k-1} \rangle_{Q_{k-1}}
  }\, .
\end{equation}
It is convenient to instead of working with different tangent spaces $T_{Q_k}\O(N)$, 
to express the search directions and gradients directly in terms of the translated in $o(N)$ arising from right multiplication with $Q_k^T$.  
Let $g_k = G_k Q^T_k$ and $h_k = H_k Q^T_k$.  
Then a quick calculation shows that the update of the search direction can be expressed as 
\begin{equation}
  h_k = - g_k + \gamma_k h_{k-1}\, .
\end{equation}
Following the proposal of Ref.~\cite{abrudan_conjugate_2009}, 
the quantity 
$\gamma_k$ can be further approximated by 
\begin{equation}
  \gamma_k = \frac{
    \langle g_k - g_{k-1}, g_k\rangle_{\Id}  
  }{
    \langle g_{k-1}, g_{k-1} \rangle_{\Id}
  }\, .
\end{equation}

Finally, we update 
\begin{align}
    Q_{k+1} = \exp(t_{k} h_k) Q_{k}, 
\end{align}
with step size $t_k$ determined by a line search algorithm {introduced in Ref.~\cite{abrudan_conjugate_2009}} that approximates the minimum of the objective function along the direction of $h_k$ with a low-order polynomial. 
Thereby, we find the optimal step size $t_k$ with only few cost function evaluations. We summarize the conjugate gradient algorithm as \cref{alg:CG}. 

\begin{figure}[tb]
\begin{algorithm}[H]
  \caption{$\operatorname{conjGrad}(f, Q_0, \epsilon)$}\label{alg:CG}
  \begin{algorithmic}[1]
    \Require Objective function $f$, initial point $Q_0 \in \O(N)$, tolerance $\epsilon$.
    \State Set $k = 0$ 
    \Repeat
        \State Calculate the Euclidean gradient $G^E_k = \nabla_E f (Q_k)$ (here using \eqref{eq:EuclideanGradient}, \eqref{eq:EuclideanGradient:part1}, and \eqref{eq:EuclideanGradient:part2}).
        \State Calculate the translated Riemannian gradient $g_k = G^E_k Q_k^T-  (G^E_k)^T Q_k \in o(N)$. 
        \State Calculate gradient norm $n_k = \langle g_k, g_k \rangle_{\Id}$
        \If{$ k = 0$}
          \State Set $h_k = - g_k$
        \Else
          \State Set $\gamma_k = (n_k - \langle g_k, g_{k-1} \rangle_{\Id}) / n_{k-1}$. 
          \State Determine search direction as $h_k = - g_k + \gamma_k h_{k-1}$
        \EndIf
        \State Perform line-search to determine $t_k$ as arg\,min of $t \mapsto f(\exp(h_k t)Q_k)$. 
        \State Set $Q_{k+1} = \exp(h_k t_k) Q_k$. 
    \Until{$n_k < \epsilon$ at $k = \hat k$}
    \Ensure objective point $Q_{\hat k}$ and objective value $f(Q_{\hat k})$. 
  \end{algorithmic}
\end{algorithm}
\end{figure}

\paragraph*{Regularization.}

Above, we have neglected the sparsity constraint.  
This has resulted in an unconstrained optimization problem over the non-convex manifold $\O(N)$.  
A straightforward way to include the model constraints on the support of the Hamiltonian term is via an additional regularization term in the objective function. 
Specifically, in the regularization we replace the optimization problem \eqref{eq:generalEigReconstruction2} by the problem
\begin{equation}\label{eq:nonConvexRegularized}
\begin{split}
  \operatorname*{minimise}_{Q \in \O(N)}&\quad f_{y,A^\lambda,\mu}(Q) \coloneqq f_{y,A^\lambda}(Q) + \mu\, r_\Omega(Q)
\end{split}\end{equation}
with the regularizer 
\begin{equation}
r_\Omega := \| (H^\lambda(\Pi(Q)))_{\bar\Omega}\|^2_F 
\end{equation}
and $\mu > 0$.  
The rationale behind the choice of the Frobenius norm as opposed to, say, the $\ell_1$-norm of the weight of $h$ on the complement of the support set---which might be the natural choice, see Ref.~\cite{HangleiterEtAl:2020}---is the following:
While the $\ell_1$-norm is the natural regularizer promoting  sparsity of a matrix, it also leads to a badly conditioned optimization problem since the gradient is non-continuous. This introduces steep edges in the optimization landscape. 
Conversely, the Frobenius norm constitutes a smooth regularizer which fares much better in the gradient descent algorithm. Moreover, since we are only interested in minimizing the total weight on the complement of the support, the corresponding regularizer effectively acts as a---slightly reweighted---$\ell_1$ norm. 

We observe that the conditioning of the optimization problem becomes worse when the data deviates from the constraint. 
This means that making a suitable choice of $\mu$ is a challenging problem. If we choose $\mu$ too large, the optimization problem becomes badly conditioned and the algorithm does not converge. 
If we choose it too small, the constraint is not enforced. 
We therefore proceed by running the optimization algorithm for increasing, exponentially spaced values of $\mu$ until it does not converge. We then perform binary search over $\mu$ to find the largest value of $\mu$ such that the algorithm converges. 


\section{Addressing SPAM errors}
\label{sec:spam}

As discussed in detail in the main text, in the realistic experimental setting the measurement data is unavoidably and significantly altered by SPAM errors. 
We can model those errors as invertible linear maps $S$ and $M$, corresponding to state preparation and measurement errors, respectively, obtaining the noisy data model (see also \cref{eq:spam time-evolution data}) 
\begin{align}
\label{eq:spam data model}
y[l] = \frac 12 M \exp(-\i t_l h) S.
\end{align}
This data model assumes that the initial and final ramp phases are to a good approximation particle number preserving.
We allow for $S,M$ to be general invertible linear maps rather than restrict them to being unitary in order to model incoherent effects during the ramping phases. 
In the presence of incoherent effects our model describes the behaviour of the dominant eigenvector of the density matrix.

In the following we outline the algorithmic strategies we use in order to alleviate the effect of $S$ and $M$ on the recovery.
We begin in \cref{ssec:sperror} by discussing the pre-processing step that removes \emph{either} the state preparation \emph{or} the measurement (SPOM) error from the data. 
In fact, this strategy enables us to fully characterize SPOM errors in the post-processing of the Hamiltonian identification.
In \cref{ssec:measurement error}, we subsequently discuss in detail the remaining error which we take---without loss of generality---to be the measurement ramp.

\subsection{Removing and characterizing SPOM errors}
\label{ssec:sperror}

To discuss SPOM errors, we consider data with \emph{either} state preparation \emph{or} measurement errors, which has the form 
\begin{align}
\label{eq:measurement error model}
    y[l] &= \frac 12 M \exp(- \i t_l h), \,\, \text{or }\\ 
\label{eq:state prep error model}
    y[l] &= \frac 12 \exp(-\i t_l h) S. 
\end{align}
Without loss of generality, here we discuss state preparation errors, i.e., data of the form \eqref{eq:state prep error model}. 
Our conclusions follow analogously for SPOM errors described by a final map $M$.

To begin with, let us write the data \eqref{eq:state prep error model} in eigendecomposition as 
\begin{align}
    y[l] &= \frac 12 \exp(-\i t_l h) S\\
    & = \frac 12 \sum_{k=1}^N \e^{- \i t_l \lambda_k} \ketbra{v_k}{v_k} S. 
    \nonumber
\end{align}
Observe that, as in the error-free case, each coefficient matrix $\ketbra{v_l}{v_l}S$ has unit rank.

{This allows us to remove the initial map from the data series in the pre-processing of the identification algorithm by forming the data series
\begin{equation}
  y^{(l_0)}[l] = y[l](y[l_0])^+ = \sum_{k=1}^N e^{-i\lambda_k(t_l - t_{l_0})} \proj{v_k}.
\end{equation}
In doing so, all entries of the data series are now affected by the noise corrupting $y[l_0]$. 
To improve noise robustness, we concatenate the data series for various $l_0$ to obtain, given integers $s,w \le L$, the extended data series
\begin{equation}\label{eq:spom removal}
    y_{\text{tot}} = (y^{(0)},y^{(s)},y^{(2s)},\dots,y^{(\lfloor L/s \rfloor s)}),
\end{equation}
where each $y^{(l_0)}[l]$ is restricted to $l \in [l_0-w, l_0+w]$.
We summarize the SPOM error removal algorithm in \cref{alg:SPERRORremoval}.

The algorithm performs $Ls^{-1}$ inversions of $N\times N$ matrices and $(2w+1)Ls^{-1}$ multiplications, requiring $\mc O(Ls^{-1}wN^3)$ flops in total. 
The resulting data series gets inflated to size $\mc O(s^{-1}wL)$.
Choosing larger values of $s$ and smaller values of $w$, thus, controls not only the time and storage complexity of the SPOM removal algorithm but also of the consecutive algorithmic steps of the identification algorithm.
}
Executing algorithm~\ref{alg:SPERRORremoval} on $y[l]^T$ instead of $y[l]$ as an input and transposing the matrices in the returned time series removes errors in the measurement instead of the state preparation.

\begin{figure}[tb]
\begin{algorithm}[H]
  \caption{$\operatorname{SPOMremoval}(y, s, w)$}\label{alg:SPERRORremoval}
  \begin{algorithmic}[1]
    \Require $y \in \CC^{(L+1)\times N \times N}$, $s \leq L$, $w \leq L$.
    \For{$l \in [\lfloor L/ s \rfloor]$}
        \State Calculate $P =  (y[ls])^{+}$.
      \For{$k \in [ls - w, ls + w]$}  
        \State Calculate $y^{(l)}[k] = y[k]\cdot P$.
      \EndFor
    \EndFor
    \Ensure the concatenation $y_{\text{tot}} = (y^{(1)}, y^{(2)}, \ldots, y^{(\lfloor L/ s \rfloor)})$.
  \end{algorithmic}
\end{algorithm}
\end{figure}

\paragraph*{SPOM error characterization.}
Using the input data $y_{\text{tot}}$, in which the initial map has been removed, in the two-step Hamiltonian reconstruction algorithm, we obtain an estimate for the Hamiltonian $\hat h$. 
We can use $\hat h$ to obtain a tomographic estimate of the initial map $S$ via
\begin{equation}
\label{eq:identified initial map}
  \hat S = \frac{2}{L+1}\sum_{l =0}^L\exp[\i t_l \hat h] y[l]\, ,
\end{equation}
or alternatively of the final map $M$ if we chose to remove it from $y_{\text{tot}}$ instead of $S$.

\subsection{The measurement error}
\label{ssec:measurement error}

Algorithm~\ref{alg:SPERRORremoval} removes either the initial or the final map from the data.
Removing the initial map still leaves us with the final ramp $M$ as a source of systematic error, the effect of which we discuss in the following.
This error in fact explains the systematic error of the method observed in the experiments presented in the main text.

When the initial map is a general invertible matrix with no further restrictions, it is impossible to uniquely identify an arbitrary final map at the same time.
This is because our model \eqref{eq:spam data model} contains a gauge freedom; a simultaneous transformation of $S,M$ that leaves the data $y[l]$ invariant.
As argued in the paragraph \emph{Imbalance between initial and final ramping phase} in Methods C, we expect the final ramp $M$ to be nearly diagonal. 
This provides us with additional structure.
Note that even if the diagonality assumption does not hold exactly, the gauge freedom allows us to partially enforce it.

The frequency estimation is robust against systematic errors due to the non-trivial final map.
The systematic errors  the identification therefore originate from the eigenspace reconstruction step.
Deriving analytical expressions for how the measurement error biases the result of the non-convexly constrained optimization problem \eqref{eq:generalEigReconstruction2} and algorithm~\ref{alg:CG} is not straight-forward. 
For this reason, we instead focus on the relaxation to the linear inversion problem \eqref{eq:linearEigReconstruction2} 
with the unique solution $\Pi = \left(A^\lambda\right)^+ Y$.
Let $Y_0, \Pi_0$ be the input data and solution to \eqref{eq:linearEigReconstruction2} in the absence of a final map.
Including $M$, the data has the form $Y = Y_0 (M^T \otimes M^{-1})$ and, thus, the solution becomes $\Pi = \Pi_0 (M^T \otimes M^{-1})$.
Note that here we have removed any initial ramp from the data using \cref{alg:SPERRORremoval}.
In \cref{alg:linInvPP}, we project the rows of $\Pi$ to real-valued unit-rank projectors, to obtain the eigenspaces $P_k$ of $\tilde h$.
In the case of a diagonal unitary $M = \diag(\e^{i\phi_1}, \ldots, \e^{i\phi_N})$, the rows of $\Pi$ are already rank-one projectors and, taking the real part, the eigenspace estimates are $P_k = C(\phi) \circ \ketbra{v_k}{v_k}$ with $\circ$ the Hadamard product (entry-wise multiplication) and 
\begin{equation}
  C(\phi) = \begin{bmatrix}
    1 & \cos(\phi_1 - \phi_2) & \dots & \cos(\phi_1 - \phi_N) \\
    \cos(\phi_1 - \phi_2) & 1 \\
    \vdots & & \ddots & \vdots \\
    \cos(\phi_1 - \phi_N) & & \dots & 1
\end{bmatrix}\,.
\end{equation}
Hence, also the recovered Hamiltonian becomes
\begin{equation}\label{eq:randDiagUnitaryM recovery}
  \tilde h = C(\phi) \circ h_0 \, .
\end{equation} 
We find that the recovery is exact on the diagonal and there is a systematic error in the sign and amplitude of the off-diagonal terms.
In \cref{ssec:systematic error} we estimate the magnitude of this error in the case where $M$ is given by a linear model of the final ramping phase using analytical and numerical evidence.
Furthermore, in \cref{sec:sign flips} we propose an algorithm that removes the sign part of the systematic error under mild assumptions on the quality of the implementation of the target Hamiltonian.

\section{The complete identification algorithm}
\label{sec:full algorithm}

\cref{alg:HamRec} summarizes the complete algorithm for Hamiltonian reconstruction.
The input to the algorithm are the data $y$,
the support of the Hamiltonian model $\Omega$ and the time grid spacing $\Delta t$ of the samples.
The step size $s$ and the window size $w$ control the SPOM removal \cref{alg:SPERRORremoval}. 
Both ESPRIT algorithms require a Hankel dimension $K$, tensorESPRIT further requires the sample set $\mc S$. 
The non-convex reconstruction using a conjugate gradient descent additionally requires a gradient tolerance $\epsilon$, an initial point $Q_0$ and a regularization parameter $\mu$.  
As demonstrated in the numerical benchmarks, $Q_0$ can be chosen at random. 
The success rate can be additionally improved by restarting the algorithm from another random initialization if the objective function is above a pre-defined threshold. 
For the recovery from experimental data, we initialize $Q_0$ at the eigenprojectors of the targeted Hamiltonian.

\Cref{alg:HamRec} as stated here is robust against errors in the state-preparation and returns an estimate for $\hat S$. 
As explained in Section~\ref{sec:spam}, we arrive at a variant of \Cref{alg:HamRec} that is robust under errors in the measurement by applying the SPOM-removal algorithm to the transposed data and estimate $\hat M$ (instead of $\hat S$) in the last step.

\subsection{Estimating the time complexity}

Let us summarise the time complexity of the individual steps of the reconstruction algorithm: 
SPOM removal takes $\mc O(s^{-1}wLN^3 + L^2N)$ flops. 
The ESPRIT algorithm on the result of the SPOM-removal step as input requires $\mc O(s^{-1}wL N^2)$ flops,
alternatively tensorESPRIT, which does not use the SPOM-removal step, requires $\mc O((LN^2 + |S|)N^3)$ flops. 
Post-projected linear inversion contributes $\mc O(s^{-1}wL N^2 + N^3)$ flops. 
Run-time estimates for the non-convex conjugate gradient algorithm are more involved as they depend on speed of convergence of the decent algorithm and of the matrix exponentiation.
We suspect that the complexity scales linearly in $s^{-1}wL$ and as a low-degree polynomial in $N$.  Thus, we expect that it is not dominating the parametric dependence of the run-time.
Roughly speaking, the quadratic `blow-up' of the data in the SPOM-removal step (assuming $w \in \mc O(L)$, $s \in \mc O(1)$)  and 
choosing $|\mathcal{S}| \in \mc O(L^2)$ in tensorESPRIT
causes all algorithmic steps to scale at most as $\mc O(L^2N^3)$.
This step also determines the required storage to be in $\mc O(s^{-1}wL N^2)$.

\begin{figure}[tb]
\begin{algorithm}[H]
  \caption{$\operatorname{HamRec}(y,\Delta t, s, w, K, [\mc S], [\epsilon, Q_0, \mu, \Omega])$}\label{alg:HamRec}
  \begin{algorithmic}[1]
    \Require 
    Data $y \in \CC^{(L+1) \times N \times N}$, 
    sample rate $1/\Delta t$,
    SPOM removal parameters $s, w$, Hankel dimension $K$, 
    [\,\emph{for} $\operatorname{tensorESPRIT}$: sample set $\mc S$\,],
    [\,\emph{for} $\operatorname{conjGrad}$: tolerance $\epsilon$, initialization $Q_0$, regularization $\mu$, support $\Omega$\,]
    \State $y_\text{total} = \operatorname{SPOMremoval}(y, s, w)$.
    \State Extract the eigenfrequencies using $z = \operatorname{ESPRIT}(\tr y_\text{total}, K ,s,w)$, or using $z = \operatorname{tensorESPRIT}(y, K, \mc S) $.
    
    \State $\lambda_k = - \Im(\log(z_k)) /(\Delta t)$ for $k \in [N]$.

    \State Calculate matrix $A^\lambda$ depending on $s, w$. 
    
    \State Reconstruct the eigenspace projectors either using $\Pi = \operatorname{linInvPP}(y_\text{total} , \A^\lambda)$ or $Q = \operatorname{conjGrad}(f_{y_\text{total},A^\lambda, \mu}, Q_0, \epsilon)$ and set $\Pi = \Pi(Q)$. 
    \State Set $\hat h = H^\lambda(\Pi)$.
    \State Set $\hat S$ according to \cref{eq:identified initial map}.
    \Ensure Hamiltonian coefficient matrix $\hat h$, initial map $\hat S$. 
  \end{algorithmic}
\end{algorithm}
\end{figure}

\subsection{Increasing the predictive power}\label{ssec:iterative procedure}

When benchmarking the performance of the algorithms in numerical simulations, where the `true' Hamiltonian generating the data is known a priori, 
it is instructive to consider not only the recovery error of the Hamiltonian itself
 but also the achieved fitting error to the data. 
 Due to the gauge freedom in the SPAM data model \eqref{eq:spam data model} different tuples $(h,S,M)$ give rise to the same observed data. 
 For this reason, the fitting error is more suitable to draw conclusions about the capabilities of an estimate $(\hat h,\hat S,\hat M)$ to predict further time series data with the same ramping phase---in a strict reading of terminology, shifting from the question of Hamiltonian identification to the question of Hamiltonian learning. 
 We refer to 
\begin{equation}\label{eq:prediction error}
  \mathcal{E}_\text{\rm pred} \coloneqq \frac1{N\sqrt{L+1}}\sum_{l=0}^L \sqrt{\|\hat y[l] - y[l]\|_{\ell_2}^2}\,,
\end{equation}
as the \emph{prediction error} in the following.

We find that we can further reduce the prediction error the recovered model exhibits by running our recovery procedure iteratively multiple times, alternatingly optimizing the initial and the final map.
After running \cref{alg:HamRec} once, {we obtain the estimate $(\hat h^{(0)},S^{(0)},M^{(0)}=\Id)$, } update the data to 
\begin{equation}
y^{(1)}[l] = (y[l] (\hat S^{(0)})^{-1})^t 
\end{equation}
and rerun the reconstruction algorithm.
We, thus, obtain a new tuple $(\hat h^{(1)},\hat S^{(1)} = \Id, \hat M^{(1)})$.
Now updating the data series to 
\begin{equation}
y^{(2)}[l] = (\hat M^{(1)})^{-1} y^{(1)}[l]
\end{equation}
and rerunning the algorithm yields a {tuple} $(\hat h^{(2)}, \hat S^{(2)}, \hat M^{(2)}=\Id)$ and so forth.
The final estimate $(\hat h, \hat S, \hat M)$ of $(h,S,M)$ can be computed from the tuples $(\hat h^{(i)},\hat S^{(i)}, \hat M^{(i)})$ for $i \in [0,\dots,r]$ via 
\begin{equation}
    \hat h = \hat h^{(r)}, \quad \hat S = \hat S^{(r)} \dots \hat S^{(0)}, \quad \hat M = \hat M^{(0)} \dots \hat M^{(r)}.
\end{equation}
We present numerical results on this iterative procedure in \cref{ssec:systematic error}.
We find that already one iteration improves the systematic prediction error significantly, while having little effect on the systematic analog implementation error.


\begin{figure}[tb]
  \centering
  \includegraphics{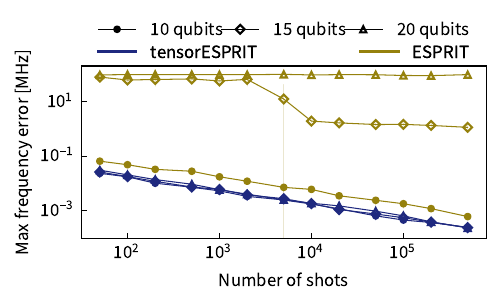}
  \caption{
    \textbf{The $\ell_\infty$-error of the recovered spectrum of tensorESPRIT (blue) and ESPRIT (mustard) with varying levels of shot noise.}
    Every point is averaged over $10$ random comb Hamiltonians with $N=10$ (circles), $15$ (diamonds) and $20$ (triangles). The error bars represent the standard deviation and are mostly smaller than the markers.
    SPOM removal pre-processing \eqref{eq:spom removal} is applied.
    All markers of tensorESPRIT coincide. 
  }
  \label{fig:esprit shot noise}
\end{figure}

\section{Numerical benchmarks}
\label{sec:numerics}

We here conduct a detailed analysis of the performance of the various stages of our algorithm on simulated data.
The Hamiltonian simulation and algorithm are implemented in the Python language.
For frequency extraction, the Hankel dimension is set to $K = \lfloor L/2 \rfloor$.  
Note that this increases the computational complexity of the algorithms compared to the optimal choice of $K \in \O(N)$.
For tensorESPRIT the set $\mc S$ is chosen according to \eqref{eq:tesprit:stdS}.
The non-convex optimization is initialized with $Q_0$ drawn at random from the Haar measure on $\O(N)$ for the numerical tests.
The success rate is additionally improved by restarting the algorithm from another random initialization if the objective function is above a pre-defined threshold. If the SPOM removal pre-processing step is applied, we use parameters $s=5$ and $w=L$.

\subsection{Data models and Hamiltonian ensembles}

We simulate the time evolution for total time $T = 0.6\,\text{\textmu s}$ with sample rate $r = 1/\Delta t = 250$\,MHz ($L = 150$) under the following ensembles of random non-interacting Hamiltonian:
\begin{itemize}
  \item \emph{Random comb Hamiltonians.} $h_{\text{comb}} = Q \diag(\lambda) Q^T$, where $\lambda = (\lambda_1, \ldots, \lambda_N)$ consists of equally spaced frequencies in the range $[-18.4, 17.0]$\,MHz and $Q$ is a Haar random orthogonal $N \times N$ matrix.
  \item \emph{Random banded Hamiltonians.} Let $\nu$ be the uniform distribution on $[0, 20]$\,MHz.
The diagonal entries $(h_{\text{banded}})_{k,k}$ are $N$ independent samples from $\nu$. The entries of the first off-diagonals $(h_{\text{banded}})_{k,k+1} = (h_{\text{banded}})_{k,k-1}$ are $N-1$ independent samples from $\nu$.
All other entries of $h_{\text{banded}}$ are zero.
  \item \emph{Random Harper Hamiltonians.} $(h_{\text{Harper}})_{k,k+1} = (h_{\text{Harper}})_{k,k-1} = -20$\,MHz and $(h_{\text{Harper}})_{k,k} = 20 \cos(2\pi k b)$\,MHz, where $b$ is drawn uniformly at random from $[0,1]$. All other $(h_{\text{Harper}})_{k,l}$ with $|k-l| > 1$ are zero.
\end{itemize}

\paragraph*{Simulating noise and errors.}
We simulate the time series data according to \eqref{eq:spam data model} with initial and final ramps $S$ and $M$ computed using one of the following prescriptions:
\begin{itemize}
  \item \emph{Random unitary.} Drawn from the Haar measure on $U(N)$.
  \item \emph{Random diagonal unitary.} $\diag(e^{i\phi_j})$, with $(\phi_j)_{j=1}^N$ i.i.d.\ samples from the uniform distribution over $[0,2\pi)$.
  \item \emph{Constant-$v$ model.} Ramp model from \cref{ssec:systematic error}. Idling frequencies of the qubits are drawn from the uniform distribution on $[-400,100]$\,MHz. For the benchmarks we set the parameters to $v = 790$\,MHz/ns and $\tau = 0.05$\,ns. 
\end{itemize}
To account for shot noise induced by a finite number $\sigma$ of samples for each expectation value, we replace each entry $y_{i,j}[l]$ by a sample from $\frac1\sigma\left(\text{B}(\sigma,\Re\{y_{i,j}[l]\}) + \i \text{B}(\sigma,\Im\{y_{i,j}[l]\})\right)$, where $\text{B}(n,p)$ is the binomial distribution for $n$ trials and probability of success $p$.

\subsection{Frequency extraction}\label{ssec:esprit benchmarks}

\begin{figure}[tb]
  \centering
  \includegraphics{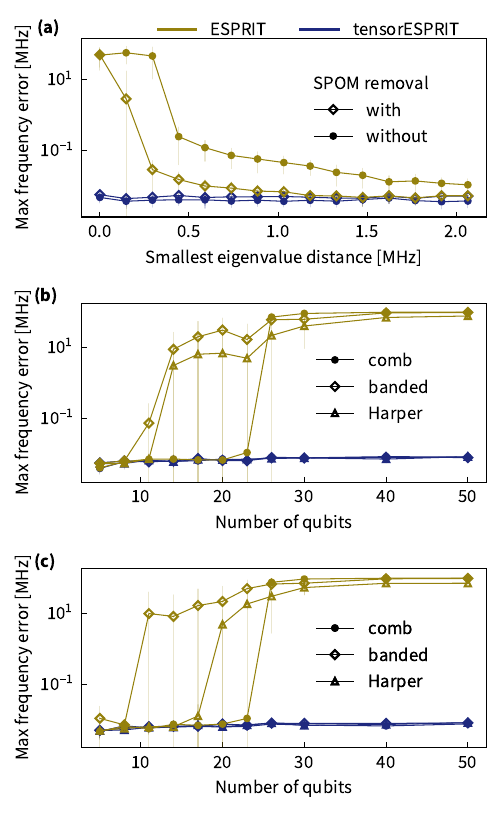}
  \caption{
      \textbf{The $\ell_\infty$ recovery errors of tensorESPRIT (blue) and ESPRIT (mustard) for (nearly) degenerate spectra and different system sizes.} 
      Shot noise with $\sigma=1000$ is applied. The error bars represent the standard deviation.
      \textbf{(a)} Recovery of the spectra of random comb Hamiltonians ($N=5$) with the second largest eigenvalue varied. Average over $30$ runs with (diamonds) and without (circles) using SPOM removal pre-processing. $x$-axis displays the distance of the second and third largest eigenvalues. $S=M=\id$.
      \textbf{(b)} Average frequency recovery error over $10$ runs of random comb, banded and Harper Hamiltonians versus the system size for $S=M=\id$ and \textbf{(c)} for Haar-random unitaries $S$ and $M$. SPOM-removal step applied.
  }
  \label{fig:esprit comparison}
\end{figure}

\begin{table}[tb]
  \begin{tabular}{l c c c r }
    \toprule
    \emph{System size} $N$ & $5$ & $20$ & $50$ & $100$\\
    \midrule
    ESPRIT [s] & $2.0 \times 10^{-2}$ & $8.8 \times 10^{-2}$ & $1.1$ & $9.9$ \\
    tensorESPRIT [s] \quad& $5.3 \times 10^{-2}$ & $7.9 \times 10^{-1}$ & $9.0$ & $81$ \\
    \bottomrule
  \end{tabular}
  \caption{\textbf{Run-times of ESPRIT and tensorESPRIT} on modern consumer grade SoC (Apple MacBook Air M1).
  Random comb Hamiltonians on various system sizes have been used to simulate $L=150$ time steps.
  The run-times of ESPRIT include the SPOM removal pre-processing step \eqref{eq:spom removal} with algorithm parameters set to $s=5,w=L$.}
  \label{tab:esprit}
\end{table}

In this section, we benchmark the ability of ESPRIT and tensorESPRIT, introduced in \cref{ssec:scalar esprit} and \cref{ssec:tensor esprit}, to recover Hamiltonian frequencies from the simulated data.
We demonstrate super-resolution capabilities of both algorithms.
We find that tensorESPRIT is capable of recovering completely degenerate spectra with no decrease in accuracy, 
making it the more scalable approach.

We first examine the dependence of the recovery on the number of shots $\sigma$ used to estimate each expectation value. 
We use both algorithms to recover the spectrum of random comb Hamiltonians with varying $\sigma$ 
for three system sizes $N=10$, $15$, and $20$.
We set $S=M=\id$. 
The SPOM removal procedure \eqref{eq:spom removal} is still used to increase the signal-to-noise ratio.
The $l_\infty$-error of the recovered frequencies with respected to the their true values averaged over 10 Hamiltonian instances is plotted in \cref{fig:esprit shot noise}.
We find that the recovery error of tensorESPRIT scales as $\sigma^{-\frac12}$.  
Using tensorESPRIT, all instances are recovered up-to shot noise limitation. 
For small system size $N=10$ the recovery error of ESPRIT shows the same scaling and recovers all instances with comparable accuracy. 
For system sizes $N=15$ and $20$, ESPRIT has a large recovery error with a phase transition appearing for $N=15$ at $\sigma = 5000$, above which some instances can be regarded as recovered. 
This can be explained by the fact that above $N=15$ the frequency spacing of random comb Hamiltonian instances (with fixed bandwidth) is too narrow to be resolved by the ESPRIT algorithm. 

To highlight this effect and demonstrate the stability of tensorESPRIT, we next examine the recovery of (nearly) degenerate spectra, with and without the effect of the SPOM removal  \eqref{eq:spom removal}.
To this end we use random comb Hamiltonians ($N=5$), where the second largest frequency is varied. 
\cref{fig:esprit comparison}, panel (a), displays the $\ell_\infty$ recovery error of both algorithms as a function of the distance between the second and the third largest frequencies. 
ESPRIT exhibits a phase transition and fails to recover spectra with small frequency spacing.
Taking a look at the output spectra of the ESPRIT algorithm, we observe that ESPRIT misses one of the nearly degenerate frequencies and substitutes it with a frequency that originates in the noise subspace.
Comparing the recovery with and without a preceding SPOM-removal step, we find that SPOM removal significantly improves the resolution capabilities of ESPRIT, {even though no SPOM error is present.}
In contrast, we see that neither the distance between the frequencies nor the SPOM removal step affect the recovery performance of tensorESPRIT.
{The effect of the SPOM-removal step on ESPRIT can be understood by the fact that this step makes all Fourier coefficients in $\tr y_\text{tot}$ equal to one (up to incoherent noise), which improves the performance of ESPRIT \cite{LiLiaoFannjiang:2019:Super-resolution}.
On the other hand it does not improve the performance of tensorESPRIT, since the pseudoinverse in forming the matrices $A^{(k,l,k',l')}$ already has similar impact.}

The remaining panels of \cref{fig:esprit comparison} show the performance in recovering the spectra of random comb, banded and Harper Hamiltonians of increasing system sizes, without (panel (b)) and with (panel (c)) SPOM errors.
ESPRIT performs well only for small system sizes. The admissible system sizes depend on the Hamiltonian ensemble. 
The recovery of tensorESPRIT is successful also for large system sizes for all three Hamiltonian ensembles. 
We do not observe a deterioration of the recovery when including Haar random unitaries as SPOM matrices $S$ and $M$ for tensorESPRIT.

In summary, we conclude that ESPRIT is suitable for recovering spectra with sufficiently well-separated frequencies as is typically found for small system sizes. 
The shortcomings of ESPRIT in resolving degenerate spectra are resolved by tensorESPRIT, demonstrating consistent performance for larger system-sizes.
This however comes at a cost of increased empirical computation times of tensorESPRIT compared to ESPRIT, \cref{tab:esprit}.

\begin{figure}
  \centering
  \includegraphics{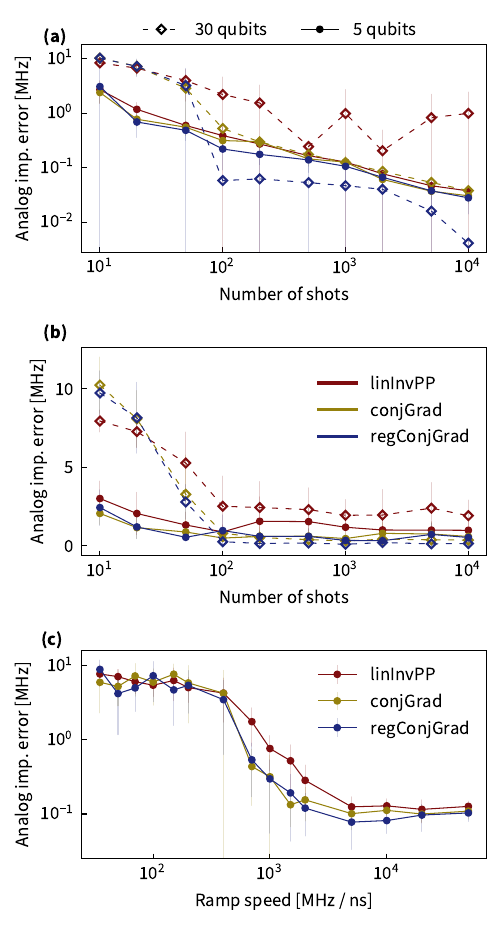}
  \caption{\textbf{Eigenspace reconstruction of random Harper Hamiltonians} of size $N=5$ (circles) and $30$ (diamonds).
  Recovery error (analog implementation error metric) averaged over $10$ instances for linear inversion with post-projection (linInvPP, red), conjugate gradient descent (conjGrad, mustard) and regularized conjugate gradient descent (regConjGrad, blue). Error bars indicate the standard deviation.
  Eigenfrequencies are extracted using tensorESPRIT, SPOM removal pre-processing \eqref{eq:spom removal} is applied.  Recovery error for 
  varying levels of shot noise \textbf{(a)} without SPAM errors ($S=M=\id$) and 
  \textbf{(b)} for $S$ a Haar-random unitary and $M$ given by the constant-$v$ model.
  \textbf{(c)} Effect of ramping speed in the constant-$v$ model on the recovery ($\sigma=1000$).
  \vspace{1.2cm}
  \label{fig:eigenspace benchmark}
  }
\end{figure}

\begin{table}[tb]
  \begin{tabular}{l l c c c c }
    \toprule
    \emph{System size $N$} & & $5$ & $10$ & $25$ & $50$\\
    \midrule
  \multirow{2}{60pt}{linInvPP [s]} & (a) \quad& $1.1$$\times$$10^{-3}$ & $4.3$$\times$$10^{-3}$ & $2.3$$\times$$10^{-2}$ & $1.4$$\times$$10^{-1}$ \\
    & (b) & $2.2$$\times$$10^{-3}$ & $4.9$$\times$$10^{-3}$ & $2.2$$\times$$10^{-2}$ & $1.8$$\times$$10^{-1}$\\
    \multirow{2}{60pt}{conjGrad [s]} & (a) & $2.4$$\times$$10^{-2}$ & $8.4$$\times$$10^{-2}$ & $2.2$ & $16.4$ \\
    & (b) & $8.7$$\times$$10^{-2}$ & $9.9$$\times$$10^{-2}$ & $1.6$ & $2.3$$\times$$10^{2}$ \\
    \multirow{2}{60pt}{regConjGrad [s]} & (a) & $4.5$$\times$$10^{-2}$ & $8.4$$\times$$10^{-2}$ & $1.8$ & $40.0$ \\
    & (b) & $1.1$$\times$$10^{-1}$ & $1.6$$\times$$10^{-1}$ & $3.1$ & $2.8$$\times$$10^2$ \\
    \bottomrule
  \end{tabular}
  \caption{\textbf{Run-times of linear inversion with post-projection (linInvPP), conjugate gradient descent (conjGrad) and regularized conjugate gradient descent (regConjGrad)} on modern consumer grade SoC (Apple MacBook Air M1).
  Simulation with random banded Hamiltonians on varying system sizes with (a) no shot noise and (b) shot noise ($\sigma=1000$). SPOM removal pre-processing \eqref{eq:spom removal} is applied.
  }
  \label{tab:optimization}
\end{table}

\subsection{Eigenspace reconstruction}
\label{ssec:eigenspace reconstruction benchmarks}

In section \cref{sec:basis reconstruction} we proposed different methods to solve the optimization problem \eqref{eq:generalEigReconstruction} in order to find the Hamiltonian eigenvectors, given eigenfrequencies recovered by ESPRIT or tensorESPRIT.
We here compare the performance of linear inversion with post-projection (linInvPP), non-convex conjugate gradient descent over $\O(N)$ (conjGrad) and regularized conjugate gradient descent over $\O(N)$ (regConjGrad).
The first two methods ignore the support constraint and solve the remaining unconstrained problem.
RegConjGrad imposes a relaxed support constraint via regularization.
In this section, we benchmark these methods, in particular the effect of structure constraints on the robustness of the protocol against statistical and systematic errors.

First, we study how the number of shots $\sigma$ used to estimate each expectation value impacts the performance of the eigenspace reconstruction methods.
To this end, we simulate the time evolution under a random Harper Hamiltonian on $5$ and $30$ qubits and add varying levels of shot noise, before running the three recovery procedures.
The distance between the recovered Hamiltonian and the one used in the simulation, the recovery error, is measured in terms of the analog implementation error \eqref{eq:implementation error} in the main text. 
The average recovery errors are displayed in \cref{fig:eigenspace benchmark} (a)
 without SPAM errors and (b) with SPAM errors included in the simulation.
All reconstruction algorithms are able to recover the $5$ qubit Hamiltonian. 
In the absence of SPAM errors, the recovery error is asymptotically compatible with a scaling as $\mc O(\sigma^{-\frac12})$.
For $N=30$ regConjGrad has significantly better recovery results than the other methods. 
The error of conjGrad and regConjGrad exhibits a phase transition around $\sigma=100$.
A similar behaviour can be seen in the presence of SPAM errors. 
We, thus, conclude that for larger systems exploiting the support constraints improves the stability.
At the same time linInvPP is most sensitive to the conditioning of the linear inverse problem for larger system sizes.
We generally observe that the convergence becomes considerably more sensitive to the instances and initial condition when including the regularization term, hinting at a more rugged optimization landscape. 
Tuning the regularization parameter as described in \cref{ssec:nonconvex} can significantly improve the convergence here.

\begin{figure}[tb]
  \centering
  \includegraphics{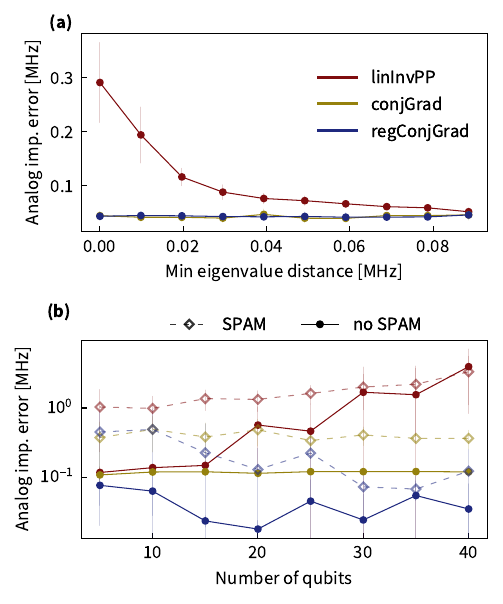}
  \caption{\textbf{Recovery error of Hamiltonians with (nearly) degenerate spectra} for linInvPP (red), conjGrad (mustard), regConjGrad (blue) averaged over $10$ instances. Error bars indicate standard deviation.
  We use tensorESPRIT for frequency extraction and apply SPOM removal pre-processing \eqref{eq:spom removal}.
  \textbf{(a)} Data simulated with random comb Hamiltonians, $N=10$, with the sixth smallest eigenvalue shifted towards the fifth smallest eigenvalue, plotting the analog implementation error against their distance.
  \textbf{(b)} Data simulated with random Harper Hamiltonians on various system sizes without SPAM errors (circles) and with Haar random unitary $S$ and $M$ given by the constant-$v$ model (diamonds).
  \label{fig:eigenspace degeneracy benchmark}}
\end{figure}

In \cref{fig:eigenspace benchmark} (c), we study the impact of SPAM errors originating from ramps with finite ramping speed on the recovery.
See \cref{ssec:systematic error} for details on the ramping model used.
We observe that only for ramp speed above roughly $1000$\,MHz/ns the recovery consistently succeeds for all algorithms while for slow rampings recovery generally fails.
Importantly, both conjugate gradient methods are more robust than linInvPP.
The experimentally observed speed in our setting is around $790$\,MHz/ns, which lies in the regime where the difference in the performance of the methods is most pronounced.

Next, we look at the impact of (near) spectral degeneracy of the Hamiltonian on the recovery in \cref{fig:eigenspace degeneracy benchmark}. 
To this end, we vary one eigenfrequency of a random comb Hamiltonian in panel (a).
We find that for small frequency spacing, linInvPP has a comparatively large recovery error. 
In contrast, both non-convex optimization methods display consistently good recovery performance also for (near) degenerate spectra. 
For the recovery of Harper Hamiltonians on systems larger than $20$ the performance of linInvPP decreases with the system size, panel (b), as here the spectrum becomes increasingly degenerate. 
Further, we see that the regularization reduces the systematic error, especially in large systems.

\Cref{tab:optimization} details empirical run-times of the three algorithms for different system sizes and simulations with and without shot noise.

\subsection{Compressed sensing capabilities}

The linear inverse problem we solve for reconstructing the eigenvectors is typically highly over-determined. 
We measure $2(L+1) N^2$ real expectation values comprising $y$ in order to infer the $N^3$ real parameters of $\Pi$.
Taking time trace data, e.g., at sample rate $250$\,MHz for $.6$\,$\mu$s the number of measurement times $2(L+1) = 2 \cdot 250 \cdot .6 = 300$ is considerably larger than the system size on current hardware with tens of sites.

Since the reconstruction algorithm additionally explicitly exploits restricted structure of the underlying signal, we expect that recovery is still possible with considerably less expectation values---following the paradigm of compressed sensing~\cite{CompressedSensingIntroCandes}, even in the regime where the linear inversion problem becomes underdetermined.

\begin{figure}[tb]
  \centering
  \includegraphics{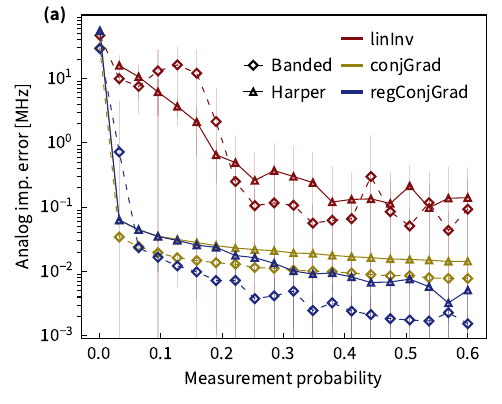}
  \includegraphics{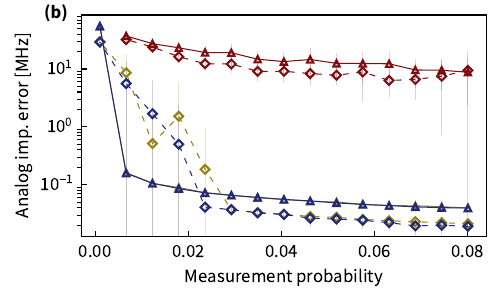}
  \caption{
       \textbf{Compressed sensing recovery.} Recovery error averaged over $30$ instances drawn from different Hamiltonian ensembles ($N=20$) and for different eigenspace reconstruction algorithms, when measurement data is randomly subsampled. Error bars display the standard deviation. \textbf{(a)} a wider range of measurement probabilities; \textbf{(b)} in the regime of an under determined inverse problem. Simulation with $\sigma = 10^6$ shots and without SPAM errors.
  }
  \label{fig:compressed sensing}
\end{figure}

We here numerically test the compressed sensing capabilities of the different algorithms.
To this end we randomly sub-select entries of $y[l]$ with a probability $p$ for the eigenvector reconstruction. 
In expectation the number of (complex) measurement settings is, thus, reduced to $2(L+1) N^2 p$.
Fig.~\ref{fig:compressed sensing} displays the recovery error for different values of $p$. 
In particular, in the regime shown in panel (b), below $p = .07$ the problem becomes underdetermined as a linear inverse problem. 
We find that conjugate gradient algorithms can successfully recover Hamiltonian instances even when the problem (without structure assumptions) is underdetermined. 
The post-projected linear inversion algorithm in contrast does only allow for moderate subsampling of the measurement entries.

\section{Estimating experimental errors}
\label{sec:error estimation}

In this section, we explain how we empirically estimate the error on the Hamiltonian $\hat h$ and initial map $\hat S$ identified via the robust identification method \cref{alg:HamRec}, including pre- and post-processing.
This error comprises two contributions. 
{First, it has a systematic contribution, which is due to the non-trivial final map $M \neq \id$ (\cref{ssec:systematic error}).
}
Second, it has a statistical contribution due to the estimation of the expectation values \cref{eq:spam data model} from finite statistics (\cref{ssec:statistical error}). 

As noted in the main text, the impact of the systematic error on the predictive power of the identified Hamiltonian is reduced by the gauge freedom in \eqref{eq:spam data model} under simultaneous transformation of $h,S,M$. 
Additionally, the systematic error in the prediction error can be further reduced using the iterative procedure described in \cref{ssec:iterative procedure}.

\subsection{Systematic error: Final ramp effect estimation}
\label{ssec:systematic error}

To estimate the magnitude of the systematic error that is induced by {a non-trivial} final pulse ramping $M$, we use an idealized model of the final ramping phase with a constant ramping speed $v$ and a padding time $\tau$ (\emph{constant-$v$ model}). 
In the experiment, ramping proceeds in three steps.
First, we ramp the coupler frequencies to turn off the hopping term in the Hamiltonian.
Second, to stabilize the frequencies, we let the system evolve under the (now diagonal) Hamiltonian for padding time $\tau$, before finally ramping the qubits to their idling frequencies to enable their measurement.
Our model assumes linear ramps with slope $v$ of all the Hamiltonian entries.
Hence, the final ramp is given by $M = U_\text{coupler} U_\text{padding} U_\text{diag}$, where 
\begin{equation}
U_j := \mathcal{T} \exp\{-i \int_0^{\tau_j} H_j(t) dt\}
\end{equation}
for $j \in \{\text{coupler},\text{diag}\}$, with $\mathcal{T}$ denoting the time-ordering operator, and $U_\text{padding} = e^{-i\tau \diag(h)}$, where $\diag$ applied to a matrix returns its restriction to the diagonal.
We set $H_\text{coupler}(t) = T_{\diag(h)}(h + \sign(\diag(h) - h) v t)$ and $H_\text{diag}(t) = T_{h_m}(\diag(h) + \sign(h_m - \diag(h)) v t)$.
Here, $h_m$ corresponds to the idle Hamiltonian at the end of the ramp pulse and the thresholding operator acts entry-wise as
\begin{align}\label{eq:final map model}
T_{g}(x)_{i,j} = \begin{cases} \min\{(g)_{i,j}, x_{i,j}\}& \text{if } \sign(h_m - h )_{i,j} > 0 ,\\
\max\{(g)_{i,j}, x_{i,j}\} &  \text{if } \sign(h_m - h )_{i,j} < 0.
\end{cases}
\end{align}
The thresholding ensures that the entries of $H(t)$ stay equal to those of $\diag(h)$ and $h_m$ respectively once they reach their final value in each ramping phase.
{The integration limits $\tau_\text{coupler}, \tau_\text{diag}$ are the minimal times at which all entries of $H_\text{coupler}(t),H_\text{diag}(t)$ reach $\diag(h), h_m$ respectively.}
We assume that the matrix after the ramp pulse $h_m$ is a diagonal matrix with frequencies corresponding to the idling frequencies of the qubits. 

Below, we empirically build trust in this model and estimate the model parameters.
Using the empirically inferred model parameters, we 
estimate the systematic errors in the following way:
Using the empirical estimates $\hat h,\hat S$ (the output of \cref{alg:HamRec}),
we simulate the time evolution using the model \eqref{eq:spam data model} with $M$ given by the constant-$v$ model. 
Running the identification \cref{alg:HamRec} again on the simulated data yields a second (bootstrapped) estimate  $\tilde h,\tilde S$. 
We use $\mathcal{E}_\text{analog}(\hat h,\tilde h)$ as an estimate for the systematic error of the analog implementation error.
By comparing $\tilde h$ with $\hat h$ entrywise, we arrive at an entrywise systematic error estimate.
Similarly, we can compute the systematic error in $\mathcal{E}_\text{analog}(\hat S,\id)$ via $\mathcal{E}_\text{analog}(\hat S,\tilde S)$.

\begin{figure}
  \includegraphics{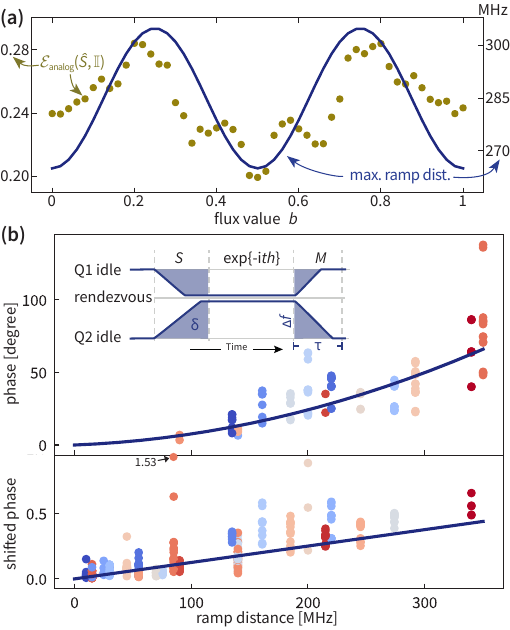}
  \caption{
  \label{fig:ramp phase} 
  \textbf{Validating the ramp model. }
  \textbf{(a)} Distance of the identified initial map before post-processing $\hat S'$ from the identity for the $5$-qubit butterfly data of \cref{fig:initial vs final} of the main text (golden dots) and maximum ramp distance $\max_i|(h_0 - h_m)_{i,i}|$ (solid line) for each flux value $b \in [0,1]$.  
  \emph{Inset illustration of the ramp model.} The qubits initially at frequencies Q1 and Q2 are ramped to the common rendezvous frequency of 6500 MHz giving rise to an initial map $S$, where they evolve under the Hamiltonian $h$ for time $t$ until they are ramped back to their idle frequencies, giving rise to a final map $M$.
  The shaded areas show the total acquired phase $\delta $ during the ramp phases. 
  {\textbf{(b)} Phases accumulated on various connected $5$-qubit subsets of the chip.
  \emph{Top.} Phase accumulated on the qubit with maximum ramp distance from each subset.
  The fit is a quadratic function with zero offset, which gives estimates $v = 800 \pm 80$\,MHz/ns, $\tau_\text{tot} = 0.09 \pm 0.03$\,ns.
  \emph{Bottom.} For the remaining qubits from each subset, shifted phase $\xi$ given by \eqref{eq:xi ramping model parameter estimation} is plotted. 
  The fit is a linear function with zero offset, which gives the estimate $v = 797 \pm 4$\,MHz/ns.}
  }
\end{figure}

\paragraph*{Empirical validation of ramp model and parameter estimation.}
Our model for estimating the systematic error induced by the final ramping phase implies that the deviation of the initial and final ramp from the identity transformation depends on the ramp distance, that is, the absolute value of the entries of $h- h_m$. 
Indeed, the maximal ramp distance is expected to set the time-scale of the ramp phase and, thus, determines magnitude of the ramping effect in the data. 
In \cref{fig:ramp phase}(a) we validate that, indeed, the deviation of the identified initial map $\hat S$ is proportional to the ramp distance $\max_{i,j}|(h- h_m)_{i,j}|$. 

In order to give an estimate of the model parameters $v$ and $\tau$, we implement the zero Hamiltonian and reconstruct it with our identification method.
In the rotating frame of the idle frequencies of the qubits, we effectively observe a diagonal Hamiltonian with eigenfrequencies that are the difference between the common rendezvous frequency (6500 MHz) and the idle frequencies.  
Since no couplers are involved, both the corresponding final and initial ramping maps are diagonal and contribute a complex phase to the data which is proportional to the combined surface area underneath the ramps, see the inset of \cref{fig:ramp phase}(b). 

\begin{figure}
  \includegraphics[width=\linewidth]{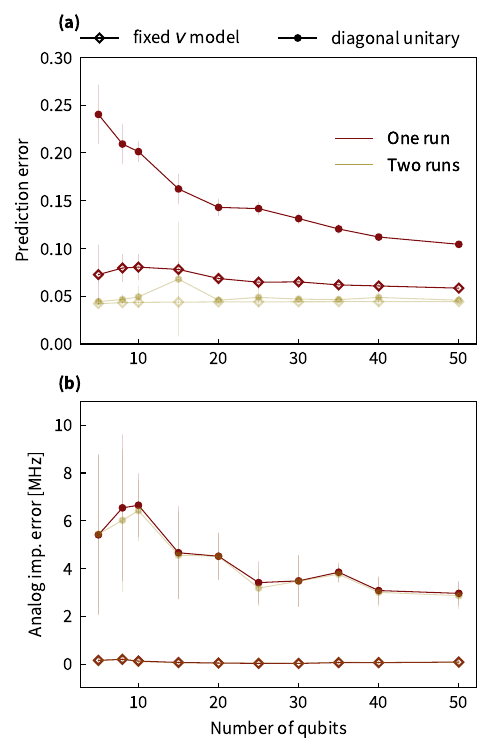}
  \caption{\textbf{Systematic recovery and prediction error due to non-trivial final map.}
  Recovery of random Harper Hamiltonians on various system sizes, using one (red) and two (mustard) runs of the algorithm (see \cref{ssec:iterative procedure} for details).
  SPAM errors modeled by a random unitary $S$ and either a random diagonal unitary $M$ (circles) or $M$ given by the constant-$v$ model (diamonds).
  The shot noise corresponds to $\sigma=1000$ shots per expectation value.
  TensorESPRIT and regularized conjugate gradient descent are used in the recovery.
  The error bars represent the standard deviation over $10$ instances.
  \textbf{(a)} Prediction error \eqref{eq:prediction error} of the recovered model. 
  \textbf{(b)} Analog implementation error of the recovered Hamiltonian.
  }
  \label{fig:final map robustness}
\end{figure}

Since the Hamiltonian is itself diagonal, it commutes with the final diagonal unitary so that, effectively, the data can be described as 
$y_{\text{diagonal}}(t) = \exp(- \i t h) MS$ with only SPOM error with an effective initial map $MS$ present. 
We can thus determine the combined phase accumulated on each qubit during the true initial and final ramping directly as the phases of the diagonal entries of our estimate $\hat S$ of the effective initial map. 
{Using the constant-$v$ model for $S$ and $M$, the magnitude of the total accumulated phase on the $j$-th qubit can be computed to be $|\phi_j| = 2\pi \Delta_j\left({\Delta_\text{max}}/{v} + \tau_\text{tot}\right)$, where $\tau_\text{tot}$ is the sum of the padding times of the initial and final ramping phases and $\Delta_j$ is the ramping distance of the $j$-th qubit.
$\Delta_\text{max}$ is the maximum ramping distance in the set of qubits involved in the experiment and it sets the ramping time for all other qubits.
In our experiment, we implemented the zero Hamiltonian on various $5$-qubit connected subsets of the chip and hence $\Delta_\text{max}$ differs for each subset.

To estimate the model parameters, in panel (b) of \cref{fig:ramp phase} we plot the magnitude of the phase accumulated on the qubit with the maximum ramping distance in the given subset against the maximum ramping distance for that subset.
Fitting a quadratic function with zero offset gives us estimates $v_1 = 800\pm 80$\,MHz/ns and $\tau_\text{tot} = 0.09 \pm 0.03$\,ns.
Now we turn to the remaining qubits.
Using the estimated padding time, we compute
\begin{equation}\label{eq:xi ramping model parameter estimation}
  \xi_j \coloneqq \frac{\frac{|\phi_j|}{2\pi} - \tau_\text{tot}}{\Delta_\text{max}} = \frac1v \Delta_j.
\end{equation}
Using a linear fit we get a second ramp speed estimate $v_2 = 797 \pm 4$\,MHz/ns, which is consistent with $v_1$ within experimental error, building trust in the model.
Note that $v=790$\,MHz/ns and $\tau=0.05$\,ns (which is the part of $\tau_\text{tot}$ we assign to the final ramp) is the parameter value used in the numerical benchmarks of \cref{sec:numerics}.
To get a conservative estimate of the systematic errors, and in face of the variance in the data \cref{fig:ramp phase}, we use $v=350$\,MHz/ns and $\tau = 0.1$\,ns.
The fit is shown in panels (b) and (c) of \cref{fig:ramp phase}, where we excluded one qubit, which accumulated seemingly random phase in each run of the experiment.

\paragraph*{Comparison to systematic error in numerical simulation.}
In order to build trust in our estimation method for the systematic error, 
we simulate the time evolution of random Harper Hamiltonians on varying system sizes with SPAM errors, and study how well the identification procedure performs in terms of analog implementation error of the recovered Hamiltonian and prediction error. 
We use random unitary initial maps and final maps that are either a random diagonal unitary or given by the constant-$v$ model.
Since the model with random diagonal unitaries can even change the sign of the recovered interaction strength, in this setting we make use of the additional post-processing step described in \cref{sec:sign flips}.
In \cref{fig:final map robustness}, panels (a), (b) and (c), respectively, we plot the prediction error, the analog implementation error and the improvement of the errors when two iterations of the method are used, in the way described in \cref{ssec:iterative procedure}.

When comparing these results to the experimental data \cref{fig:single H,fig:increasing system size,fig:initial vs final} in the main text, 
we see that the prediction errors are close to the prediction errors achieved by the constant-$v$ model. 
This constitutes an independent validation of our method of estimating the systematic error in the Hamiltonian identification.
In panel (c), we can see that the iterative procedure significantly improves the systematic error in the prediction error, while having little effect on the systematic error on the analog implementation error.

\paragraph*{Scaling estimate of the systematic error.}
In panel (b) we observe that the systematic error decreases with the system size for the random diagonal unitary model for $M$.
Even though not apparent in the parameter regime of the numerical benchmarking, we expect a similar dependence also for the constant-$v$ model. 
This can be understood using the following theoretical reasoning.
We have seen in \cref{ssec:measurement error} that in the absence of statistical errors and assuming a diagonal unitary $M=\diag(e^{i\phi_1},\dots,e^{i\phi_N})$, the recovered Hamiltonian is given by \eqref{eq:randDiagUnitaryM recovery}. 
Suppose the case of a one-banded Hamiltonian with couplings of typical magnitude $\bar J$, where the diagonal entries have typical magnitude $\bar h$ and need to be ramped by a typical distance $\bar \Delta$. 
If we assume the linear ramping model above with constant ramping speed $v$ and a padding time $\tau$, we can estimate the magnitude of the systematic error in the analog implementation error\footnote{Here we are assuming that all the qubits are ramped in the same direction, which is close to reality, and neglect the ramping of the couplers.}  to be
\begin{equation}\label{eq:systematic error estimate}
  \mathcal{E}_{\text{syst}} \approx \bar J (2/N)^{\frac12}\left[1-\cos\left(\pi\left(\frac{\Delta^2}{v} + 2\tau \bar h\right)\right)\right],
\end{equation}
which using realistic values $\Delta = 300$\,MHz, $\bar h = \bar J = 20$\,MHz, $v =350$\,HMz/ns, $\tau = 0.1$\,ns evaluates to approximately ${6.4\sqrt2}/{\sqrt{N}}$\, MHz, showing that the systematic error should decrease with system size for one-banded Hamiltonians, in accordance to \cref{fig:final map robustness}. 
Note that this argument itself relies on relaxations of the original problem, assumptions on the final map $M$ and the experimental characterization of the model. 
On this basis it further serves as a consistency check for the magnitude of the reported systematic errors. 
}

\paragraph*{Remaining sign-errors.} 
As reported in the main text, we observe that the Hamiltonian identification algorithm recovers 
some interactions with the opposite sign than the target Hamiltonian. 
Such an error can be explained by the presence of a diagonal unitary final map with phase differences $\pi/2$ and $3\pi/2$. 
However, we find that the constant-$v$ ramp model with the empirically estimated parameters and in the regime of the observed prediction error does never produce final maps with such large phase differences. 
Furthermore, the random diagonal ramp model that produces matrices $M$ with sufficient phase differences to flip signs in the recovered interactions yields prediction errors that are considerably larger than the ones we observe in the experiment, see \cref{fig:final map robustness}. 
The observed sign flips, thus, point to a separate source of sign systematic error. 
Under the assumption that sign flips originate from SPAM, we can however efficiently correct for them in the post-processing. 
We explain the corresponding post-processing algorithm for reconstructing $M$ restricted to an orthogonal diagonal rotation in \cref{sec:sign flips}.

\subsection{Statistical error: Bootstrapping}
\label{ssec:statistical error}

Let us now turn to estimating the statistical error of the identification result. 
We estimate the size of the error induced on the Hamiltonian estimate that is returned by the identification method via parametric bootstrapping. 
To this end, we simulate time series data with finite statistical noise according to the model \eqref{eq:spam data model} with $M = \id$ using the identified Hamiltonian $\hat h$ and a Haar-random unitary for the initial ramp $S$.  
We then run the Hamiltonian identification method with conjGrad on $10^5$ instances of such synthetic data. 
As the statistical error of the entry we use the $0.99$-quantile ($99\%$ confidence level) of the absolute deviation of each entry in the Hamiltonians obtained from the synthetic data, from the corresponding entry of the identified Hamiltonian used to generate the data. 
We observe that the statistical errors of the entries are of comparable size and only report the maximal statistical error over all entries. 

We also calculate $0.99$-quantile of the deviation of the synthetically identified Hamiltonian $\hat h_{\text{bt}}$ from the originally identified Hamiltionian $\hat h$ in terms of the analog implementation error 
$\mathcal{E}_{\rm analog}(\hat h_{\text{bt}},\hat h)$, (see \cref{eq:implementation error} of the main text), and likewise for the eigenfrequencies.  
This is used as the statistical error estimate for the analog implementation error benchmark.

Omitting the regularization in the identification method reduces the computational complexity of the bootstrapping and produces more well-behaved empirical distributions of the deviation error.  
At the same time the regularization is shown in \cref{ssec:eigenspace reconstruction benchmarks} to improve the estimate and, thus, the statistical error obtained in this way is expected to dominate the statistical error of the regularized identification method. \smallskip

\begin{figure}[tb]
  \begin{algorithm}[H]
    \caption{OneBandedCorrectFlips($h, h_0$)}\label{alg:one banded final map recovery}
    \begin{algorithmic}[1]
      \Require symmetric $h \in \RR^{N \times N}$, symmetric one-banded $h_0 \in \RR^{N \times N}$.
      \State Set $D_M = \id$.
      \State Define $G_\pm(A) = A h A^T \pm h_0$.        \For{$m \in \{2,\dots,N\}$}
          \State Set $a_0 = |G_-(D_M)[m-1,m]|$.
          \State Set $a_{\text{flip}} = |G_+(D_M)[m-1,m]|$.
          \If{$a_{\text{flip}} < a_0$}
            \State Set $(D_M)[m,m] = -1$.
          \EndIf
        \EndFor
      \State Set $\hat h = D_M h D_M$.
      \Ensure sign-fixed Hamiltonian coefficient matrix $\hat h$, final map estimate $D_M$.
    \end{algorithmic}
  \end{algorithm}
\end{figure}

\begin{figure}[tb]
  \begin{algorithm}[H]
    \caption{GreedyCorrectFlips($h, h_0$)}\label{alg:final map recovery}
    \begin{algorithmic}[1]
      \Require symmetric $h \in \RR^{N \times N}$, symmetric $h_0 \in \RR^{N \times N}$.
      \For{$m\!\neq\!n$ in the order of decreasing $|h[m,n]|$, s.t. neither $m$ nor $n$ has been probed previously}
        \State Set $a_0 = |G_-(D_M)[m,n]|$.
        \State Set $a_{\text{flip}} = |G_+(D_M)[m,n]|$.
        \If{$a_0 < a_{\text{flipped}}$}
          \State Set $D_M^{\text{flip}} = D_M$.
          \State Set $D_M^{\text{flip}}[m,m] = D_M^{\text{flip}}[n,n] = -1$.
          \If{$\|G_-(D_M^{\text{flip}})\|_F < \|G_-(D_M)\|_F$}
            \State Set $D_M[m,m] = D_M[n,n] = -1$.
          \EndIf
        \Else
          \State Set $D_M^{1} = D_M^2 = D_M$.
          \State Set $D_M^{1}[m,m] = -1$, $D_M^2[n,n] = -1$.
          \If{$\|G_-(D_M^1)\|_F < \|G_-(D_M^2)\|$}
            \State Set $D_M[m,m] = -1$.
          \Else 
            \State Set $D_M[n,n] = -1$.
          \EndIf
        \EndIf
      \EndFor
      \State Set $\hat h = D_M h D_M$.
      \Ensure sign-fixed Hamiltonian coefficient matrix $\hat h$, final map estimate $D_M$.
    \end{algorithmic}
  \end{algorithm}
\end{figure}

\section{Reconstructing diagonal orthogonal final maps}\label{sec:sign flips}
We partially remove the systematic error induced by the final ramping phase with the following post-processing procedure.
Suppose we are given the estimates $\hat h, \hat S$ by the identification \cref{alg:HamRec}. 
To remove the sign part of the systematic error on the Hamiltonian recovery, we determine an orthogonal diagonal $\hat M = \hat D_M$ that solves
\begin{equation}\label{eq:final map estimation}
  \operatorname*{minimize}_{D_M = \diag{\pm 1}}\quad \| D_M \hat h D_M - h_0 \|_F^2
\end{equation}
and perform the gauge transformation
\begin{equation}\label{eq:spam fixing gauge transformation}
  \begin{split}
    \hat h' = \hat D_M \hat h \hat D_M, \\
    \hat S' = \hat D_M \hat S,
  \end{split}
\end{equation}
to obtain the model $(\hat h', \hat S', \hat D_M)$. 
This estimate further reduces the systematic errors compared to $(\tilde h, \tilde S, \id)$.

Note that in general solving \eqref{eq:final map estimation} is an NP-hard problem. 
To see this, consider the case when $h_0$ is a matrix with all ones and $\hat h$ is a matrix with entries $\pm 1$.  
The problem then encodes the maximum balanced subgraph problem, which is known to be NP-hard \cite{bartholdiGoodSubmatrixHard1982}.
However, for 1D nearest-neighbour hopping Hamiltonians, where the Hamiltonian coefficient matrices are one-banded, we can give an efficient algorithm:
The algorithm sets $\hat D_M[1,1] = +1$ and then updates the diagonal entries of $\hat D_M$ one-by-one.
For the element $\hat D_M[m,m]$ with $m>1$, it decides to set it to $\pm1$, picking in each turn the option more favourable to the cost function. 
Since in each of the $N-1$ steps we take into account one more independent off-diagonal element of $h_0$, of which there are also exactly $N-1$, this procedure finds the exact solution in the one-banded case. The algorithm is summarized in \cref{alg:one banded final map recovery}.

For general Hamiltonian, one can apply the following greedy heuristic algorithm:
We create a list of off-diagonal elements of $\tilde h$, ordered according to their decreasing absolute value. 
Then, we set $\hat D_M = \id$ and loop over the entries of the list.
At each step, the algorithm decides whether flipping the sign of the element $h[m,n]$ under consideration would decrease the cost function.
If yes, it sets either $\hat D_M[m,m] = -1$ or $\hat D_M[n,n] = -1$, depending on which one is better with respect to the cost function.
If not, it either does nothing or sets both $\hat D_M[m,m] = \hat D_M[n,n] = -1$, depending on which option is better with respect to the cost function.
The algorithm is summarized in \cref{alg:final map recovery}.
We note that for the problem sizes we encounter in the experiment we can also exactly solve the minimization problem \eqref{eq:final map estimation} through exhaustive search.  


\putbib
\end{bibunit}

\end{document}